\documentclass[twoside,8pt]{article}
\usepackage{epsfig}

\newcommand{\be}{\begin{equation}}
\newcommand{\ee}{\end{equation}}
\newcommand{\bea}{\begin{eqnarray}}
\newcommand{\eea}{\end{eqnarray}}

\begin{document}

\title{Nuclear symmetry energy effects on neutron stars properties }

\author{V.P. Psonis, Ch.C. Moustakidis and S.E. Massen  \\
$^{}$ Department of Theoretical Physics, Aristotle University of
Thessaloniki, \\ 54124 Thessaloniki, Greece }

\maketitle

\begin{abstract}
We construct a class of nuclear equations of state based on a
schematic potential model, that originates from the work of
Prakash et. al. \cite{Prakash-88}, which reproduce the results of
most microscopic calculations. The equations of state are used as
input for solving the Tolman-Oppenheimer-Volkov  equations for
corresponding neutron stars. The potential part contribution of
the symmetry energy to the total energy is parameterized in a
generalized form both for low and high values of the baryon
density. Special attention is devoted to the construction of the
symmetry energy in order to reproduce the
results of most microscopic calculations of dense nuclear
matter. The obtained nuclear equations of state are applied for
the systematic study of the global properties of a neutron star
(masses, radii and composition). The calculated masses and radii of
the neutron stars are plotted as a function of the potential part parameters of
the symmetry energy. A linear relation between these parameters, the radius and
the maximum mass
of the neutron star is obtained. In addition, a linear relation between the
radius and the derivative of the symmetry energy near the
saturation density is found. We also address on the problem of
the existence of correlation between the pressure near the
saturation density and the radius.

\end{abstract}
\

\section{Introduction}

Neutron stars (NS) are some of the densest manifestations of
massive objects in the universe which provide very rich
information for testing theories of dense matter physics and also
provide a connection among nuclear physics, particle physics,
statistical physics and astrophysics
\cite{Lattimer-04,Shapiro-83,Glendenning-96,Weber,Prakash-94,Heiselberg-00,
Steiner-05,Prakash-97,Lattimer-00,Sahakian74}. The global aspects
of neutron stars, such as the masses, radii and composition are
determined by solving the so-called Tolman-Oppenheimer-Volkov
(TOV) equations \cite{Tolman-39,Volkov-39}. However there are large
variations in predicted radii and maximum masses because of the
uncertainties in the nuclear equation of state (EOS) near and
mainly above the saturation density $n_s$
\cite{Steiner-05,Dieperink-03,Dieperink-05,Stone-02,Stone-03,Stone-07,
Shetty-05,Shetty-06,Klahn-06,Danielewicz,Danielewicz-06,Danielewicz-04,Wiringa-88,Li-05,Zuo-02,Douchin-01,Baldo-97,Lee-98,
Liu-04,Krastev-06,Li-06,Li-06-2,Li-04,Vretenar-03}. The total energy of neutron
rich matter (the case of a neutron star) can be written as a sum
of two parts. The first one is the contribution of the symmetric
nuclear matter (which is well known) and the second is the
symmetry energy (SE) which still is uncertain although several
constraints exist from ground state masses (binding energies) and
giant dipole resonances of laboratory nuclei. A consequence of
this uncertainty is that different models predict up to a factor
of 6, variations in the pressure of neutron star matter near
$n_s$, even though the pressure of symmetric matter is better
known, being nearly zero at the same density. This pressure
variation accounts for the nearly 50\% variation in the
predictions of neutron star radii \cite{Lattimer-04,Lattimer-01}.

In general, the value of the SE at nuclear saturation density and
mainly the density dependence of the SE are both difficult to be
determined in the laboratory. The motivation of the present work is
to propose a new parameterization for the potential part of the
symmetry energy $E_{sym}(n)$ in order to be able to reproduce the
results of a variety of microscopic models both in low and high
values of the baryon density. Especially the trend of the symmetry energy
just above the equilibrium density $n_s$ is a critical factor in
determining the neutron star radius.

In order to calculate the global properties of neutron stars
(mass, radius ets.) the hydrostatic equilibrium
equations of Tolman, Oppenheimer and Volkov have to be solved once
the equation of state is specified. However, the composition of a
neutron star still remains uncertain and the construction of the
EOS, which is based on the ingredients of the NS's and the kind of
interactions which characterize them, is subjected to several
assumptions. In any case the calculated EOS has to satisfy the
following requirements \cite{Baldo-97}: i) It must display the
correct saturation point for symmetric nuclear matter (SNM); ii)
it must give a SE compatible with nuclear phenomenology especially at high
densities; iii) for SNM the incompressibility at saturation must be
compatible with the values extracted from
phenomenology; iv) both for neutron matter and SNM the speed of
sound must not exceed the speed of light (causality condition), at
least up to the relevant densities.

In the present work we consider that the neutron star core is
composed only by an uncharged mixture of neurons, protons and
electrons in equilibrium with respect to the weak interaction
($\beta$-stable matter). However in general, in the range of
densities $n \geq n_s$, the hadronic phase of superdense matter
with a rich spectrum of particles (hyperons, baryonic resonances,
$\pi^-$ and $K^-$ mesons, and a small portion of leptons) is
realized. The model which is used for the construction of the EOS
is a generalization of a schematic potential model based on a
previous work of Prakash et al. \cite{Prakash-88}. The model
reproduces the results of most microscopic calculations of dense
matter \cite{Wiringa-88}. It is worthwhile to notice that there
many ways to determine the equation of state through the many-body
approach of interacting hadrons. Some of the most recent ones are
based on variational methods \cite{Akmal97,Akmal98,Morales02} and
some are based on microscopic calculations \cite{Zhou04}. In order
to face the problem that stems from the uncertain behavior of the
SE at high densities we perform a suitable parameterization both
in low and high densities. In the previous work of Prakash et al
\cite{Prakash-88,Prakash-97} the parameterization of the potential
term of the SE is achieved by the introduction of three different
choices of the potential contribution to the SE. To advance, in
the present work, we suggest a more generalized parameterization
of the potential term of the SE which is more flexible and
efficient, reproduces the predictions of most microscopic
calculations of dense matter \cite{Wiringa-88} and confirms the
results of various empirical data.

The most striking feature of the proposed parameterization is the
different form of the parameterization function $F(u)$ for
densities below saturation point and for densities above this
point. This is not surprising since this was already entailed in
microscopic calculations. Although  the behavior of the SE for
densities below the saturation point still remains unknown,
significant progress has been made only most recently in
constraining the SE at subnormal densities and around the normal
density from the isospin diffusion data in heavy-ion collisions
\cite{Wen05, Bao05}. This has led to a significantly more refined
constraint on neutron-skin thickness of heavy nuclei
\cite{Steiner05, Chen05} and the mass-radius correlation of
neutron stars \cite{Li-06,Li-06-2,Li-04}. For densities above the saturation
point the trend of the SE is model dependent and exhibits
completely different behavior.

The above characteristic of the SE is well reflected in our
proposed models. In view of the previous comment, the proposed
parameterization of the potential term of the SE has the advantage
to be able to reproduce microscopic calculations in cases where
the SE, at low densities, increases along with the density and
then begins to fall although the density continues to increase.
This is a well known characteristic of a class of Skyrme
interactions \cite{Stone-02,Stone-03,Stone-07} and of Gogny Hartree-Fock
calculations \cite{Li-06,Li-06-2,Li-04}. Special effort has been devoted to find
analytical relations between the radius $R$ and the pressure $P$
which correspond to a special density $n$ for a fixed value of the
mass $M$ of the neutron star. So an accurate determination of a
neutron star radius will permit evaluation of the pressure of
neutron star matter. All the above will provide a direct
determination of the density dependence of the nuclear SE at these
densities \cite{Lattimer-01,Lattimer-03}.

Finally, we also address  on the problem of neutron star
cooling \cite{Pethick-92,Lattimer-91}. It is well known that the
direct Urca process can occur in neutron stars if the proton
concentration exceeds some critical value in the range of 11-15 \%.
The proton concentration can be  determined by the trend of the SE
especially just above the equilibrium density. So, the detailed
knowledge of the SE behavior is crucial for the existence of the
direct Urca process.

The plan of the paper is as follows. In section $2$ the proposed
model and the relatives formulas are discussed and analyzed.
Results are reported and discussed in section $3$, while the
summary of the work is given in section $4$.
\section{The model}
In general, the energy per baryon of neutron-rich matter may be
written as
\begin{equation}
\frac{E(n,x)}{A}=\frac{E(n,\frac{1}{2})}{A}+(1-2x)^2
E_{sym}^{(2)}(n)+(1-2x)^4 E_{sym}^{(4)}(n)+ \cdots \ ,
\label{En-1-general}
\end{equation}
To a good approximation, it is sufficient to retain in the above
expansion only the quadratic term. Thus, the above equation takes
the form
\begin{equation}
\frac{E(n,x)}{A}=\frac{E(n,\frac{1}{2})}{A}+(1-2x)^2
E_{sym}^{(2)}(n) \ , \label{En-1}
\end{equation}
where $n$ is  the baryon density ($n=n_n+n_p$) and  $x$ is the
proton fraction ($x=n_p/n$). The symmetry energy
$E_{sym}(n)=E_{sym}^{(2)}(n)$ can be expressed in terms of the
difference of the energy per baryon between neutron ($x=0$) and
symmetry ($x=1/2$) matter
\begin{equation}
E_{sym}(n)=\frac{E(n,0)}{A}-\frac{E(n,\frac{1}{2})}{A} \ .
\label{Esym-1}
\end{equation}

In the present work we consider a schematic equation for symmetric
nuclear matter energy (energy per baryon $E/A$ or equivalently the
energy density per nuclear density $\epsilon/n$) which is
given by the expression \cite{Prakash-88}
\begin{equation}
\frac{E(n,1/2)}{A}=\frac{\epsilon_{sym}}{n}=m_Nc^2
+\frac{3}{5}E_F^0 u^{2/3}+V(u), \qquad u=n/n_s \label{ESNM-1}
\end{equation}
where $E_F^0=(\hbar k_F^0)^2/2m_N$ is the Fermi
energy per baryon in equilibrium state and $n_s$ is the saturation
density.

The density dependent potential energy per nucleon $V(u)$ of the
symmetric nuclear matter is parameterized, based on the previous
work of Prakash et. al. \cite{Prakash-88,Prakash-97} as follows
\begin{equation}
V(u)=\frac{1}{2}Au+\frac{B u^{\sigma}}{1+B' u^{\sigma -1}}+3
\sum_{i=1,2}C_i \left(\frac{\Lambda_i}{p_F^0}\right)^3
\left(\frac{p_F}{\Lambda_i}-\arctan\frac{p_F}{\Lambda_i}\right),
\label{Vu-1}
\end{equation}
where $p_F$ is the Fermi momentum, related to $p_F^0$ by
$p_F=p_F^0u^{1/3}$. The parameters $\Lambda_1$ and $\Lambda_2$
parameterize the finite forces between nucleons. The values
used here are $\Lambda_1=1.5 p_F^0$ and
$\Lambda_2=3 p_F^0$. The parameters $A$, $B$, $B'$, $\sigma$,
$C_1$ and $C_2$ are determined with the constraints provided by
the properties of nuclear matter saturation. In the present work
the values of the above parameters are determined in order that
$E(n=n_s)/A-m_Nc^2=-16$ MeV, $n_s=0.16$ fm$^{-3}$ and $K_0=240$
MeV. In general the parameter values for three possible values of
the compression modulus $K_0$
$\left(K_0=9n_0^2\frac{d^2(E/A)}{dn^2}|_{n_0} \right)$ are
displayed  in table I, on Ref. \cite{Prakash-88}.

To a very good approximation, the nuclear
symmetry energy $E_{sym}$ can be parameterized as follows
\cite{Prakash-94}
\begin{equation}
E_{sym}(u)=\left(2^{2/3}-1\right)\frac{3}{5}E_F^0
\left(u^{2/3}-F(u)\right) +S_0 F(u), \label{Esym-2}
\end{equation}
where $S_0$ is the SE at the saturation point, $S_0=E_{sym}(u=1)$.
In general, theoretical predictions give $S_0=25-35$ MeV. In the
present work we consider $S_0=30$ MeV. The function $F(u)$
parameterizes the potential contribution of the nuclear SE and has
to satisfy the constraints $F(u=0)=0$ and $F(u=1)=1$. Equation
(\ref{Esym-2}) can be written in a more instructive form by separating
the kinetic and the potential
contribution of the SE.
\begin{equation}
E_{sym}(u) \simeq \underbrace{13 u^{2/3}}_{Kinetic}+\underbrace{17
F(u)}_{Potential}.\label{Esym-3}
\end{equation}

In the previous work of Prakash et. al. \cite{Prakash-88}, three
representative forms that mimic the results of most microscopic
models are used and have the following form
\begin{equation}
F(u)=u, \qquad F(u)=\frac{2u^2}{1+u}, \qquad F(u)=\sqrt{u}.
\label{Fu-form}
\end{equation}

In the present work we generalize the previous form of the
function $F(u)$ in  two ways. First, the function $F(u)$ is
parameterized as follows
\begin{equation}
F(u)=u^c ,\label{fu-1}
\end{equation}
where the parameter $c$ (hereafter called potential parameter)
varies between $0.4<c<1.5$ in order to get reasonable values
for the SE. It is obvious that according to the above formula the
trend of the potential part is the same both in  low and high
values of the baryon density.

The information gained from microscopic theoretical calculations
shows that this is not the general case for the potential part of
the SE. On the contrary, the SE exhibits different trends in low
and high densities. So, one should try to find a new formula for
the function $F(u)$ which satisfies the above restrictions. In the
spirit of the previous statement we propose a new parameterization
of the function $F(u)$. The new function which is more flexible
compared to the previous ones, reproduces the SE for most
realistic calculations and has the following form
\begin{eqnarray}
F(u)&=&\left\{ \begin{array}{ll}
u^{c_1}               &    \mbox{$u \leq 1$} \\
\\
u^{c_2} e^{1-u}+(u-1)(c_1+1-c_2)   &  \mbox{$u \geq 1$} \ .
                              \end{array}
                       \right.
\label{FU-2}
\end{eqnarray}

The function $F(u)$ satisfies the constraints $F(u \rightarrow
1^+)=F(u \rightarrow 1^-)$ and $F'(u \rightarrow 1^+)=F'(u \rightarrow
1^-)$. The derivative of the  function, compared to
equation (\ref{fu-1}), is determined  by the parameters $c_1$
and $c_2$ (hereafter called potential parameters).

In order to construct the nuclear equation of state, the
expression of the pressure is needed. In general, the pressure, at
temperature $T=0$, is given by the expression
\begin{equation}
P=n^2 \frac{d(\epsilon/n)}{d n}=n \frac{d\epsilon}{d n}-\epsilon .
\label{Pres-1}
\end{equation}

From equations (\ref{En-1}), (\ref{ESNM-1}) and (\ref{Pres-1}) we
found that the contribution of the baryon to the total pressure is
given by the relation
\begin{equation}
P_b=\left[\frac{2}{5}E_F^0 n_s u^{5/3}+u^2n_s \frac{d V(u)}{du}
\right] + n_s(1-2x)^2 u^2 \frac{d E_{sym}(u)}{d u} .
\end{equation}

The leptons (electrons and muons) originating
for the condition of the beta stable matter  contribute also to the total energy and total pressure \cite{Prakash-94}.
To be more precise the electrons and the muons which are the
ingredients of the neutron star are considered as non-interacting
Fermi gases. In that case their contribution to the total energy and
pressure  is given by
\begin{equation}
\epsilon_{e^-,\mu^-}=\frac{m_l^4 c^5}{8 \pi^2
\hbar^3}\left[(2z^3+z)(1+z^2)^{1/2}-\sinh^{-1}(z)\right],
  \label{lepton-en}
\end{equation}

\begin{equation}
P_{e^-,\mu^-}=\frac{m_l^4 c^5}{24 \pi^2
\hbar^3}\left[(2z^3-3z)(1+z^2)^{1/2}+3\sinh^{-1}(z)\right],
\label{lepton-pres}
\end{equation}
where $z=k_F/m_lc$. Now the total energy and pressure of charge
neutral and chemically equilibrium nuclear matter is
\begin{equation}
\epsilon_{tot}=\epsilon_b+\sum_{l=e^{-},\mu^{-}}\epsilon_l,
\label{tot-e}
\end{equation}

\begin{equation}
P_{tot}=P_b+\sum_{l=e^{-},\mu^{-}} P_l \ .\label{tot-pr}
\end{equation}

From equations (\ref{tot-e}) and (\ref{tot-pr}) we can construct
the equation of state in the form $\epsilon=\epsilon(P)$. What
 remains is the determination of the proton fraction $x$ in
$\beta$-stable matter. In that case we have the process
\begin{equation}
n \longrightarrow p+e^{-}+\bar{\nu}_e \qquad \qquad p +e^{-}
\longrightarrow n+ \nu_e
\end{equation}
that takes place simultaneously. We assume that neutrinos generated in
these reactions have left the system. This implies that
\begin{equation}
\hat{\mu}=\mu_n-\mu_p=\mu_e ,\label{chem-1}
\end{equation}
where $\mu_n,\mu_p$ and $\mu_e$ are the chemical potential of the neutron, proton and electron respectivelly.
Given the total energy density $\epsilon \equiv
\epsilon(n_n,n_p)$, the neutron and proton chemical potential
can be defined as
\begin{equation}
\mu_n=\frac{\partial \epsilon}{\partial n_n}|_{n_p}, \qquad \qquad
\mu_p=\frac{\partial \epsilon }{\partial n_p}|_{n_n} .
\label{chem-2}
\end{equation}

It is easy to show that after some algebra we get
\begin{equation}
\hat{\mu}=\mu_n-\mu_p=-\frac{\partial \epsilon /n}{\partial
x}|_n=\frac{\partial E}{\partial x}|_n . \label{chem-3}
\end{equation}

In $\beta$ equilibrium one has
\begin{equation}
\frac{\partial E}{\partial
x}=\frac{\partial}{\partial x}\left(E_b(n,x)+E_e(x)\right)=0 ,
\label{b-equil-1}
\end{equation}
where $E_b(n,x)$ the energy per baryon and $E_e(x)$ the electron
energy. The charge condition implies that $n_e=n_p=nx$ or
$k_{F_e}=k_{F_p}$. Combining the relations (\ref{En-1}) and
(\ref{chem-3}) we get
\begin{equation}
\hat{\mu}=4(1-2x)E_{sym}(n) . \label{chem-4}
\end{equation}

Finally by combining equations (\ref{chem-1}) and (\ref{chem-4})
we arrive at the relation
\begin{equation}
4(1-2x)E_{sym}(n)=\hbar c(3 \pi^2 n_e)^{1/3}=\hbar c(3 \pi^2 n
x)^{1/3} , \label{b-equil-2}
\end{equation}
where we considered that the chemical potential of the electron is
given by the relation $\mu_e=\sqrt{k_{F_e}^2c^2+m_e^2c^4}\approx
k_{F_e} c $ (relativistic electrons). Equation (\ref{b-equil-2})
determines the equilibrium proton fraction $x(n)$ once the density
dependent symmetry energy $E_{sym}(n)$ is known. After straightforward
algebra we get
\begin{equation}
x(n)=\frac{1}{2}-\frac{1}{4}\left([2\beta
(\gamma-1)]^{1/3}-[2\beta (\gamma+1)]^{1/3}\right) , \label{prf-1}
\end{equation}
where \[ \beta=3\pi^2 n(\hbar c/4E_{sym}(n))^3 ,\qquad \qquad
\gamma =\left(1+\frac{2 \beta}{27} \right)^{1/2}. \]

When the electrons energy is large enough (i.e. greater than the muon
mass), it is energetically favorable for the electrons to convert
to muons
\begin{equation}
e^{-} \longrightarrow \mu^{-}+\bar{\nu}_{\mu}+\nu_e .
\end{equation}

However, in the present work we will not include the muon
case to the total equation of state since the muon contribution
does not alter significantly the gross properties of the neutron
stars.

%
%

It is worthwhile to notice that the present model satisfies the relativistic causality. That means the
speed of sound which was defined from the relation,
\begin{equation}
\left(\frac{c_s}{c_l}
\right)^2=\frac{dP}{d\epsilon}=\frac{dP/dn}{d\epsilon/dn} ,
\label{causality}
\end{equation}
does not exceed the speed of light for any value of the baryon
density. This is a basic treat for any realistic EOS, regardless
the details of the interactions among matter constituents or the
many body approach \cite{Baldo-97}.

The most efficient process, which leads to a fast cooling of a
neutron star, is the direct Urca process involving nucleons
\begin{equation}
n \longrightarrow p+e^{-}+\bar{\nu}_e, \qquad \qquad p +e^{-}
\longrightarrow n+ \nu_e \label{beta-11} \ .
\end{equation}

This process is only permitted if energy and momentum can be
simultaneously conserved \cite{Pethick-92,Lattimer-91}. This
requires that the proton fraction must be $x>1/9 \simeq 0.11$.
From equation (\ref{prf-1}) it is obvious that the proton fraction
$x$ is sensitive to the density dependence of the SE and as
consequence to the parameterization of the potential part of
$E_{sym}(n)$. So, in the present work it is worthwhile to study
the relation of the proton fraction and the relative
parameterization and also to check if our parameterization
satisfies the constraints for the beggining of the Urca process.

In order to calculate the gross properties of a NS we assume that
a NS has a spherically symmetric distribution of mass in
hydrostatic equilibrium and is extremely cold ($T=0$). Effects of
rotations and magnetic fields are neglected and the equilibrium
configurations are obtained by solving the
Tolman-Oppenheimer-Volkoff equations \cite{Tolman-39,Volkov-39}
\begin{eqnarray}
\frac{dP(r)}{dr}&=&-\frac{G m(r) \rho(r)}{r^2}
\left(1+\frac{P(r)}{c^2 \rho(r)}\right) \left(1+\frac{4 \pi r^3
P(r)}{c^2 m(r)} \right) \left(1-\frac{2 G m(r)}{c^2 r}
\right)^{-1} , \nonumber \\
 \nonumber \\
  \frac{d M(r)}{dr}&=&4 \pi r^2 \rho(r)=\frac{4 \pi r^2
\epsilon(r)}{c^2} .
 \label{TOV-1}
\end{eqnarray}

To solve the set of equations (\ref{TOV-1}) for $P(r)$ and $M(r)$
one can integrate outwards from the origin ($r=0$) to the point
$r=R$ where the pressure  becomes zero. This point defines R as
the coordinate radius of the star. To do this, one needs an
initial value of the pressure at $r=0$, called $P_c=P(r=0)$. The
radius $R$ and the total mass of the star, $M\equiv M(R)$, depend
on the value of $P_c$. To be able to perform the integration, one
also needs to know the energy density $\epsilon(r)$ (or the
density mass $\rho(r)$) in terms of the pressure $P(r)$. This
relationship is the equation of state  for neutron star matter and
in the present work has been calculated for various cases by using
our model. It should also be noted that besides the stellar radius
and mass, other global attributes of a neutron star are
potentially observable, including the moment of inertia and the
binding energy \cite{Lattimer-01, Grigoryan83}. Thus, it would be
of interest to study the nuclear symmetry dependence on these
attributes. Such work is in progress.

\section{Results and discussion}
First we apply our model in a simple case where the potential
part of the SE is parameterized as $F(u)=u^c$ and the total SE
contribution can be written as follows
\begin{equation}
E_{sym}(u)=13u^{2/3}+17u^c .\label{Esym-c-simpler}
\end{equation}

The potential parameter $c$ varies between $0.4\leq c \leq 1.5$ which
gives reliable values of the SE. The total pressure of the cold
beta-stable nucleonic matter is given by
\begin{equation}
P(n,x)=n^2\left[\frac{E'(n,\frac{1}{2})}{A}+E_{sym}'(n)(1-2x)^2\right]+P_{e^-}(n)
, \label{Pres-tot}
\end{equation}
where $E_{sym}'(n)$ takes the form
\begin{equation}
E_{sym}'(n)\equiv \frac{d E_{sym}(n)}{d n} =\frac{1}{n_s}
\left[\frac{26}{3}u^{-1/3}+17cu^{c-1}\right] . \label{Esym-Fuc}
\end{equation}

We are interested for the total pressure at the saturation density
$n_s$. Considering that the electron pressure $P_{e^-}(n)$ is
\cite{Lattimer-01}
\begin{equation}
P_{e^-}(n)\cong nx(1-2x)E_{sym}(n) \label{Pres-electron}
\end{equation}
then the total pressure at $n_s$ is given by the expression
\cite{Lattimer-01}
\begin{equation}
P_{s}(n_s,x_{s})=n_{s}(1-2x_{s})\left[n_{s}E_{sym}'(n_s)(1-2x_{s})+E_{sym}(n_{s})x_{s}\right]
, \label{Pres-sat}
\end{equation}
where $E_{sym}'(n_s)=\left[\frac{d E_{sym}(n)}{d
n}\right]_{n=n_s}$ and the equilibrium proton fraction at $n_s$ is
given
\begin{equation}
x_s \simeq (3 \pi^2 n_s)^{-1}(4E_{sym}(n_s)/\hbar c)^3 \simeq 0.04
. \label{pr-fr-ns}
\end{equation}

For small values of $x_s$ we find that
\begin{equation}
P_s(n_s,x_s)\simeq n_s^2 E_{sym}'(n_s) . \label{Pres-sat-2}
\end{equation}

From the former expression it is obvious that the pressure is
mostly sensitive to the density dependence of the SE at the
saturation point $n_s$. Using our model from equation
(\ref{Esym-Fuc}) we get
\begin{equation}
E_{sym}'(n_s)\simeq  \frac{1}{n_s}\left(\frac{26}{3}+17c\right) .
\label{Esym-sat}
\end{equation}

From equations (\ref{Pres-sat-2}) and (\ref{Esym-sat}) it is
concluded that the relation between the pressure $P_s$ and the
potential parameter $c$ is
\begin{equation}
P_{s}(n_s,x_{s})\simeq n_s \left(\frac{26}{3}+17c\right) .
\label{Pre-c1}
\end{equation}

In order to calculate the global properties of the neutron star,
radius and mass we solved numerically the TOV equations
(\ref{TOV-1}) with the given equations of state constructed with
the present model. For very low densities ($n<0.08$ fm$^{-3}$) we used the
equation of state taken from Feynman, Metropolis and Teller
\cite{Feynman-49} and also from Baym, Bethe and Sutherland
\cite{Baym-71}.

In order to illustrate the density dependence trend of the SE
proposed in our model we display in figure 1a $E_{sym}(n)$ as
a function of the density $n$ for various values of the potential
parameter $c$. It is obvious that the parameter $c$ affects
decisively the trend of the SE, especially at high values of the
density. So, it is very interesting to study how the values of the
parameter $c$, and consequently the potential contribution,
affect the gross properties of the NS. In the same figure, results of Ref. \cite{Akmal98}
(the case A18+$\delta$u+UIX$^{*}$, see TABLE VI and VII of Ref. \cite{Akmal98}) are included.
It is found that the use of the phenomenological equation (\ref{Esym-3})
with proper value ($c=0.9$) of function (\ref{fu-1}) reproduces
the results of the above microscopic calculations.

Figure 2a demonstrates the linear dependence between the radius
$R_{max}$ and the parameter $c$. From our analysis it is concluded
that there is a direct relation between $R_{max}$ and the
parameterization of the potential part of the SE.
 The star symbol corresponds to the case  A18+$\delta$u+UIX$^{*}$ with $c_1=0.9$. The
corresponding relation was derived with the least-squares fit
method and has the form
\begin{equation}
R_{max}=9.10195+2.08304c .\label{Rmax-c-fit}
\end{equation}

In addition, in figure 2b we indicate the behavior of the neutron
star radius $R_{1.4}$, which corresponds to a neutron star mass
$M=1.4 M_{\odot}$, versus the parameter $c$.
 The star symbol corresponds to the case  A18+$\delta$u+UIX$^{*}$ with $c_1=0.9$. It is obvious that
there is also a linear relation between $R_{1.4}$ and $c$ which is
\begin{equation}
R_{1.4}=10.52114+3.56746c . \label{R14-c-fit}
\end{equation}

Figure 2c displays the correlation between the maximum mass of the
neutron star $M_{max}$ and the parameter $c$.
The star symbol corresponds to the case  A18+$\delta$u+UIX$^{*}$ with $c_1=0.9$. We found an almost
linear relation between $M_{max}$ and $c$ which has the form
\begin{equation}
M_{max}=1.87442+0.12344c . \label{Mmax-c-fit}
\end{equation}

By combining equations (\ref{Esym-sat}) and (\ref{R14-c-fit}) we
found a linear relation between $R_{1.4}$ and $E'_{sym}(n_s)$
which has the form
\begin{equation}
R_{1.4}=8.702+0.0336E'_{sym}(n_s) , \label{R14-Esym-fit-ns}
\end{equation}
and vice-versa
\begin{equation}
E'_{sym}(n_s)=-259.185+29.783R_{1.4} . \label{Esym-R14-fit-ns}
\end{equation}

In order to illustrate further the relation between the
radius $R_{1.4 }$ and the trend of the SE we plot in figure 3a
the radius $R_{1.4 }$ versus the derivative of
the symmetry energy $E'_{sym}(3n_s/2)$ at the baryon density
$n=3n_s/2$. A linear relation is found, which has the form
\begin{equation}
R_{1.4}=10.1798+0.0242E'_{sym}(3n_s/2) . \label{R14-Esym-fit}
\end{equation}

It is concluded that there is a direct relation between the radius
$R_{1.4 }$ and the trend of the SE, close to the saturation point
$n_s$. In addition in figure 3b we plot $E'_{sym}(3n_s/2)$ versus
$R_{1.4 }$ with the linear correlation
\begin{equation}
E'_{sym}(3n_s/2)=-420.40895+41.3066R_{1.4} . \label{Esym-R14-fit}
\end{equation}

To ensure the relativistic causality in the present model we
display figure 4a, where the ratio of the speed of sound to
speed of light $c_s/c_l$ is plotted versus the baryon density for
various values of the parameter $c$. Evidently, in all cases the
speed of sound does not exceed that of light even at high values
of the baryon density. In addition in figure 4b we plot the proton
fraction $x_p$ calculated from expression (\ref{prf-1}) as a
function of the baryon density $n$. It is obvious that only in the
cases where $c>0.5$ the proton fraction, after a specific density,
exceed the critical value $x^{Urca}\simeq0.11$ which ensures the
beginning of the Urca process.

We also tried to find the correlation between the pressure $P$
(and consequently the radius R) and the SE for other values of the
density $n$. In order to clarify the problem of the expected
relation between the radius  and the pressure we present a more
simplified model of a non-relativistic equation with a polytrope
type of EOS. Thus the EOS has the form
\cite{Shapiro-83,Lattimer-01}
\begin{equation}
P=K \rho^{\gamma} , \qquad \gamma=1+\frac{1}{\lambda} .
\label{EOS-Polyt}
\end{equation}
and the  radius of the star is given by
\begin{equation}
R=\left[\frac{(\lambda+1)K}{4 \pi G}
\right]^{1/2}\rho_c^{(1-\lambda)/2\lambda} \xi_1 , \label{Rad-L-E}
\end{equation}
where $\rho_c$ is the central density and $\xi_1$ is the solution
of the equation $\theta(\xi_1)=0$, where the function
$\theta(\xi)$ is the solution of the differential Lane-Emden
equation
\begin{equation}
\frac{1}{\xi^2}\frac{d}{d \xi}\xi^2\frac{d \theta}{d
\xi}=-\theta^{\lambda} . \label{L-E}
\end{equation}

Now it is obvious from equations (\ref{EOS-Polyt}) and
(\ref{Rad-L-E}) that in case $\lambda=1$ (or $\gamma=2$) where
$\xi=\pi$ we have
\begin{equation}
\frac{R}{P^{1/2}}=\left[\frac{2 G}{\pi}\right]^{1/2}\frac{1}{\rho}
\ . \label{R-P-LE}
\end{equation}

Thus, from equation (\ref{R-P-LE}) we concluded that in the case of a
polytrope with $\gamma=2$ there is a universal relation of the
ratio $R/P^{1/2}$ calculated for a specific value of the density
$\rho$. However if general relativity effects are
included in the above analysis the exponent $1/2$ of the pressure
is found to be smaller \cite{Lattimer-01}.

Following the above statement we plot in figure 5a the quantity
$R_{1.4}P^{-a}$ as a function of the radius $R_{1.4}$. One can see
that there is a correlation between the radii $R_{1.4}$ and the
pressure evaluated at densities $1n_s$, $3n_s/2$ and $2n_s$. The
values of the parameters $a$ and $C(n)$ have been defined by
least-squares fit of the expression $R_{1.4}=C(n)P^{a}$.

In addition, in figure 5b we plot the quantity $R_{1.4}R^{-1/4}$ as
a function of the $R_{1.4}$ to compare it with the previous work
of Lattimer et. al \cite{Lattimer-01}. It is worthwhile to notice
that the quantity $R_{1.4}R^{-1/4}$ is a mild increasing function
of the radius $R_{1.4}$. This effect is more evident for densities
far from the saturation( $n=3n_s/2,2n_s$). So, from our study it is
concluded that there is a slight dependence of the
quantity $R_{1.4}R^{-1/4}$ from the potential parameter $c$ and
consequently from the trend of the SE.

We proceed now in the more complicated case where the function
$F(u)$ is given by expression (\ref{FU-2}). In that case the
derivative of the $E_{sym}(n)$ is given by
\begin{eqnarray}
E_{sym}'(n)&=&\left\{ \begin{array}{ll}
\frac{1}{n_s}\left[\frac{26}{3}u^{-1/3}+17c_1u^{c_1-1}\right] &    \mbox{$u \leq 1$} \\
\\
\frac{1}{n_s}\left[ \frac{26}{3}u^{-1/3}+17\left(c_1+1-c_2+
e^{1-u}u^{c_2}(\frac{c_2}{u}-1)\right) \right] & \mbox{$u \geq 1$}
\ .
                              \end{array}
                       \right.
\label{DESYM-2}
\end{eqnarray}

The potential parameters $c_1$ and $c_2$ varied between $0.5
\leq c_1 \leq 1.2 $ and $0 \leq c_2 \leq 2$ in order
to get a reliable density dependent SE.

In figure 1b we display $E_{sym}(n)$ as a function of the density
$n$ for various values of the potential parameters $c_1$ and
$c_2$. In general the case is as follows, for fixed values of
the parameter $c_2$, the SE is an increasing function of $c_1$. In
addition, for fixed values of the parameter $c_1$ the increase of
the parameter $c_2$ leads to a decrease of the SE. It is seen
that, within the present model, the stiff or soft behavior of
$E_{sym}(n)$ found in various microscopic calculations, is
reproduced. As a comparison, similar to figure 1a, results of
Ref. \cite{Akmal98} (the case A18+$\delta$u+UIX$^{*}$) are included.
It is found that the use of the phenomenological equation (\ref{Esym-3})
with proper values ($c_1=0.77$ and $c_2=1.09$) of function (\ref{FU-2}) reproduces
the results of the above microscopic calculations.

Figure 6a illustrates the behavior of the radius $R_{1.4}$ as a
function of the second potential parameter $c_2$ for various
values of the first potential parameter $c_1$. The calculated points for
various values of $c_1$ can be reproduced by a second order polynomial.

\begin{eqnarray}
R_{1.4}&=&12.58956+0.05378c_2-0.46172c_2^2, \qquad c_1=0.5
\nonumber \\
R_{1.4}&=&13.04786-0.02752c_2-0.32340c_2^2, \qquad c_1=0.7 \nonumber \\
R_{1.4}&=&13.85705-0.08087c_2-0.21393c_2^2, \qquad c_1=1.0
\nonumber
\\
R_{1.4}&=&14.61946-0.19441c_2-0.14536c_2^2, \qquad c_1=1.2 \ .
\label{R-c1-c2-a}
\end{eqnarray}

In all examined cases, the radius $R_{1.4}$ is a decreasing
function of the potential parameter $c_2$. This is a direct
consequence of the softening of the equation of state due to
increase of the parameter $c_2$.

In addition, in figure 6b the behavior of the radius $R_{1.4}$ as
a function of the first potential parameter $c_1$ is reproduced
for various values of the second potential parameter $c_2$. The
least-squares fit values are given for the following linear
equations
\begin{eqnarray}
R_{1.4}&=&11.09603+2.85172c_1, \qquad c_2=0.0 \nonumber \\
R_{1.4}&=&11.01810+2.85517c_1, \qquad c_2=0.5 \nonumber \\
R_{1.4}&=&10.42034+3.06724c_1, \qquad c_2=1.2 \ .
 \label{R-c1-c2}
\end{eqnarray}

Unlike the previous case, $R_{1.4}$ is an
increasing function of the potential parameter $c_1$. The
increase of the parameter $c_1$ leads to the stiffness of the SE as
indicated in figure 6b. It is worthwhile to note that the slopes
of the best fit lines are almost the same and there is just a
shift of the lines depending on the values of the parameter $c_2$.

Also, from figure 6a and 6b we conclude that the radius
$R_{1.4}$ depends mainly on the parameter $c_1$ which determines
the derivative of the $E_{sym}(n)$ and also the pressure $P_{sat}$
at the saturation density $n_s$. However, there is a small
dependence on the parameter $c_2$ which is connected with the trend
of $E_{sym}(n)$ at higher values  of the density $n_s$.  Figures 6c and 6d demonstrate
the dependence of the radius $R_{max}$ and the mass $M_{max}$ respectively on the parameter
$c_1$ (for fixed values of the parameter $c_2$). The star symbol corresponds the the case
A18+$\delta$u+UIX$^{*}$ ($c_1=0.77$ and $c_2=1.09$). In both cases
a linear relations holds between $R_{max}$, $M_{max}$ and the parameter $c_1$.
The lines correspond to the least-squares fit values.

 It is of interest to compare the NS properties ($R_{1.4}$, $R_{max}$ and $M_{max}$) which
originated from the use of equations  (\ref{fu-1}) and (\ref{FU-2}). As an example we use the
parameterization of equations  (\ref{fu-1}) and (\ref{FU-2}) which reproduce very well the
$E_{sym}$ trend of the case A18+$\delta$u+UIX$^{*}$. As a result it is found that
$R_{1.4}(c=0.9)=13.42$ and $R_{1.4}(c_1=0.77, c_2=1.09)=12.76$ (difference ~5 \%) ,
$R_{max}(c=0.9)=11$ and $R_{max}(c_1=0.77, c_2=1.09)=10.59$ (difference ~3.7 \%),
$M_{max}(c=0.9)=1.978$ and $M_{max}(c_1=0.77, c_2=1.09)=1.974$ (difference ~0.2 \%). It is obvious
that in the case  A18+$\delta$u+UIX$^{*}$ we receive almost identical results for $M_{max}$ while there is
a small difference for $R_{1.4}$ and $R_{max}$. The differentiation of the values of $R_{1.4}$
is a consequence of the linear relation
which hold between the radius $R_{1.4}$ and the
derivative of the $E_{sym}$ (and consequently according to (\ref{Pres-sat-2}) to the pressure),
close to the saturation point (see figures $3$ and $7$).
More specifically, we receive for the two cases, $P_s(c=0.9)=3.83467$ MeV fm$^{-3}$ and
$P_s(c_1=0.77,c_2=1.09)=3.48107$ MeV fm$^{-3}$.
In general, the small differentiation on the radii  is not surprising since
the trend of equations  (\ref{fu-1}) and (\ref{FU-2}),
due to suitable parametrization, are similar. However, it is worth to point out
that equation  (\ref{FU-2}) is a generalization of equation (\ref{fu-1}), in the
meaning that while equation (\ref{fu-1}) describes well the case where $E_{sym}$
is a increasing function of the density, equation (\ref{FU-2}) is sufficiently
flexible  to describe
in addition the case where $E_{sym}$ at low densities increases along with the density
and then begins to decreases  although the density continues to increases.

To illustrate further  this point, we studied  the correlations
between the derivative of the symmetry energy $E'_{sym}$ and the
radius $R_{1.4}$ close to the saturation point $n=3n_s/2$. In
figure 7a we plot the radius $R_{1.4}$ versus the derivative of
the symmetry energy $E'_{sym}(3n_s/2)$ for fixed values of the
potential parameter $c_2$. One can see that there is a linear
relation between $R_{1.4}$ and $E'_{sym}(3n_s/2)$ just like in
figure 3a. The effect of the parameter $c_2$ is to induce a
parallel shift of the best fit lines. In figure 7b we indicate the
inverse relation, that means $E'_{sym}(3n_s/2)$ versus $R_{1.4}$.
The least-squares fit values for both cases and for various values of
the parameter $c_2$ are given for the following
equations
\begin{eqnarray}
R_{1.4}&=&8.70394+0.02684 E_{sym}'(3n_s/2), \qquad c_2=0 \nonumber \\
E_{sym}'(3n_s/2)&=&-318.5026+36.82989 R_{1.4} \ , \label{R-Esym-0}
\end{eqnarray}
\begin{eqnarray}
R_{1.4}&=&9.73291+0.02687 E_{sym}'(3n_s/2), \qquad c_2=0.5 \nonumber \\
E_{sym}'(3n_s/2)&=&-359.07613+36.98138 R_{1.4} \ ,
 \label{R-Esym-05}
\end{eqnarray}
\begin{eqnarray}
R_{1.4}&=&10.27305+0.02887 E_{sym}'(3n_s/2), \qquad c_2=1.2 \nonumber \\
E_{sym}'(3n_s/2)&=&-354.58213+34.54212 R_{1.4}  \ .
\label{R-Esym-12}
\end{eqnarray}

In figure 8a, likewise with the figure 4a we display the ratio
$c_s/c_l$ as a function of the density for various cases. It is
obvious that the relativistic causality is satisfied once again.
In figure 8b we display the proton fraction $x_p$ as a function of
the density for various cases. A more systematic study of the
$x_p$ leads to the conclusion that the potential parameter $c_1$
plays the most critical role for the occurence of the Urca
process. Specifically a higher value of the $c_1$ leads to the
beggining Urca process in smaller values of the baryon density.

Figure 9 illustrates the behavior of the quantity $R_{1.4}P^{-a}$
as a function of the radii $R_{1.4}$ for pressure determined at
$n=n_s$, $3n_s/2$, $2n_s$, and also for $c_2=0$ (figure 9a),
$c_2=0.5$ (figure 9b) and $c_2=1.2$ (figure 9c).

In figure 10 we plot the quantity $R_{1.4}P^{-1/4}$ as a function
of $R_{1.4}$ for the pressure determined at $n=n_s$, $3n_s/2$,
$2n_s$ and for $c_2=0$ (figure 10a), $c_2=0.5$ (figure 10b) and
$c_2=1.2$ (figure 10c). It is obvious once again that the quantity
$R_{1.4}P^{-1/4}$ is almost constant only when the pressure is
calculated at the saturation point $n_s$. When the pressure is
calculated at densities $n=3n_s/2$ and $n=2n_s$ the quantity
$R_{1.4}P^{-1/4}$ is an increasing function of the radius
$R_{1.4}$. Thus, as in the case of the simple parameterization of
the SE, it is concluded that there is a dependence of the quantity
$R_{1.4}P^{-1/4}$ from the first potential parameter $c_1$ as well
as from the second potential parameter $c_2$ and consequently from
the trend of the SE both for low and high values of the baryon
density.

\section{Summary}
In the present work we performed a systematic study of the effect of the potential part of the SE on the global properties of
neutron stars (masses, radii and composition). The potential part
of the SE was parameterized in a generalized form both for low and
high values of the baryon density in order to be efficient in reproducing the
results of most microscopic calculations of dense nuclear
matter.

In the case of the simple parameterization of the SE the most
striking feature of our study was the derivation of a linear
relation which stands between the maximum mass $M_{max}$, the
radius $R_{max}$ and the radius $R_{1.4}$ with the potential
parameter c. In addition, a linear relation stands
between the $R_{1.4}$ and the derivative of $E'_{sym}(n)$ for densities close to
the saturation point ($n=n_s,3n_s/2$). It was concluded that quantity $R_{1.4}P^{-a}$ (with $a$ and $C(n)$ fitting parameters) appears to be constant for the densities $n=n_s, 3n_s/2, 2n_s$. However, the quantity
$R_{1.4}P^{-1/4}$ exhibits an increasing behavior as a function of
$R_{1.4}$ for $n=3n_s/2,n=2n_s$.

In the case of the more complicated parameterization, where the SE is
parameterized in a different way for low and high values of
the density, similar results were taken. Specifically
$R_{1.4}$ is a function of both potential
parameters $c_1$ and $c_2$. This means that the value of
$R_{1.4}$ is affected from the density dependent trend of the SE,
both in low and high densities. However, we showed that for fixed values of the
parameter $c_2$, close to the
saturation point ($n=3n_s/2$), a linear relation  between the
$R_{1.4}$ and the $E'_{sym}(3n_s/2)$ stands. Finally, the quantity $R_{1.4}P^{-a}=C(n)$
appears to be constant after a suitable parameterization of the
parameters $a$ and $C(n)$ but still remains dependent from  the second
potential parameter $c_2$. The quantity $R_{1.4}P^{-1/4}$, as in
the previous case, exhibits an increasing behavior as a function of
the $R_{1.4}$ for density values above the
saturation point.

\section*{Acknowledgments}
The work was supported by the Pythagoras II Research project
(80861) of E$\Pi$EAEK and the European Union. One of the authors
(Ch.C. M) would like to thank Dr. Maddapa Prakash for kindly
providing his lectures {\it The Equation of State and Neutron
Star} which were delivered at a Winter School held in Puri India.


\newpage

\begin{figure}
 \includegraphics[height=8.0cm,width=8.0cm]{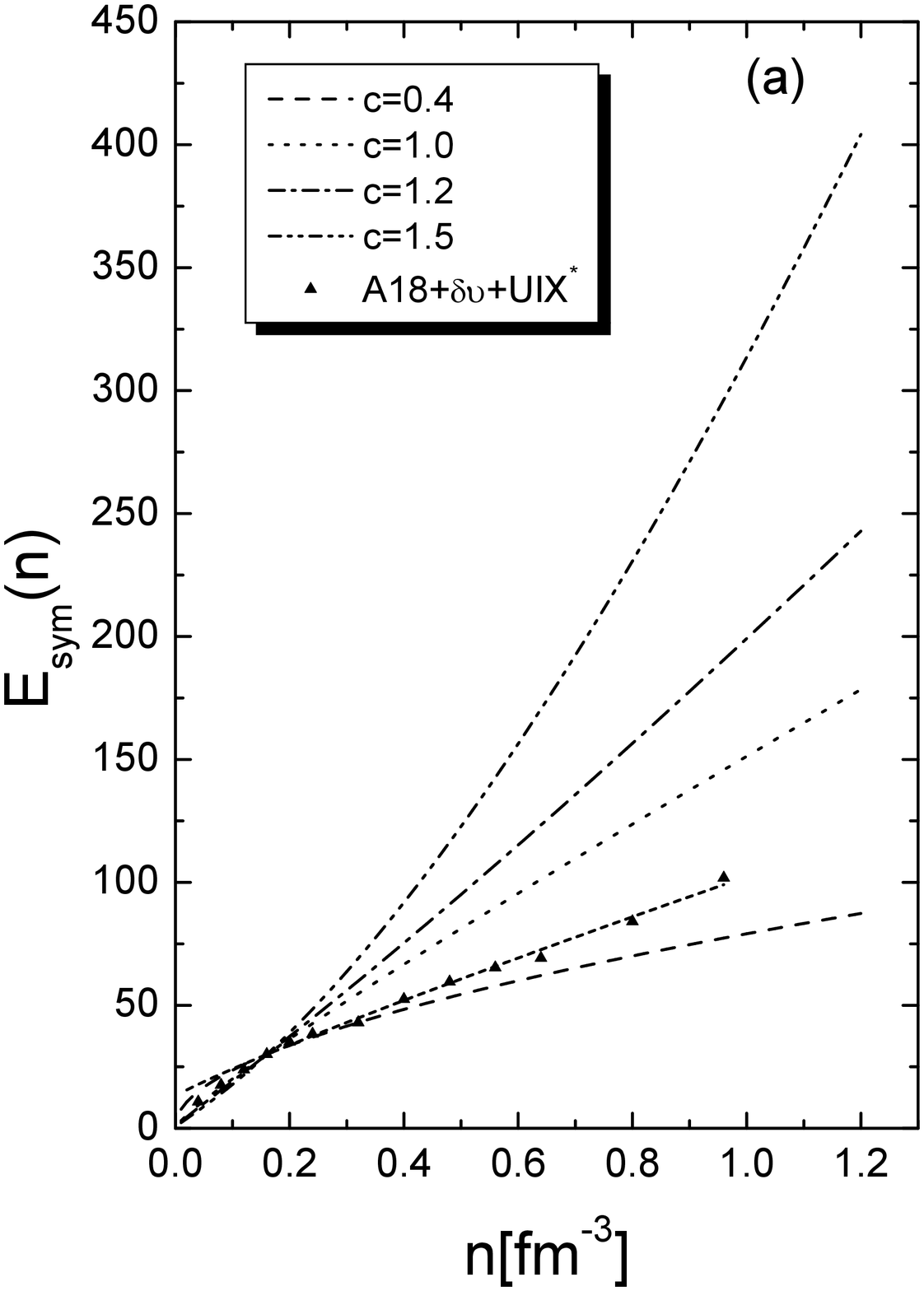}
\
 \includegraphics[height=8.0cm,width=8.0cm]{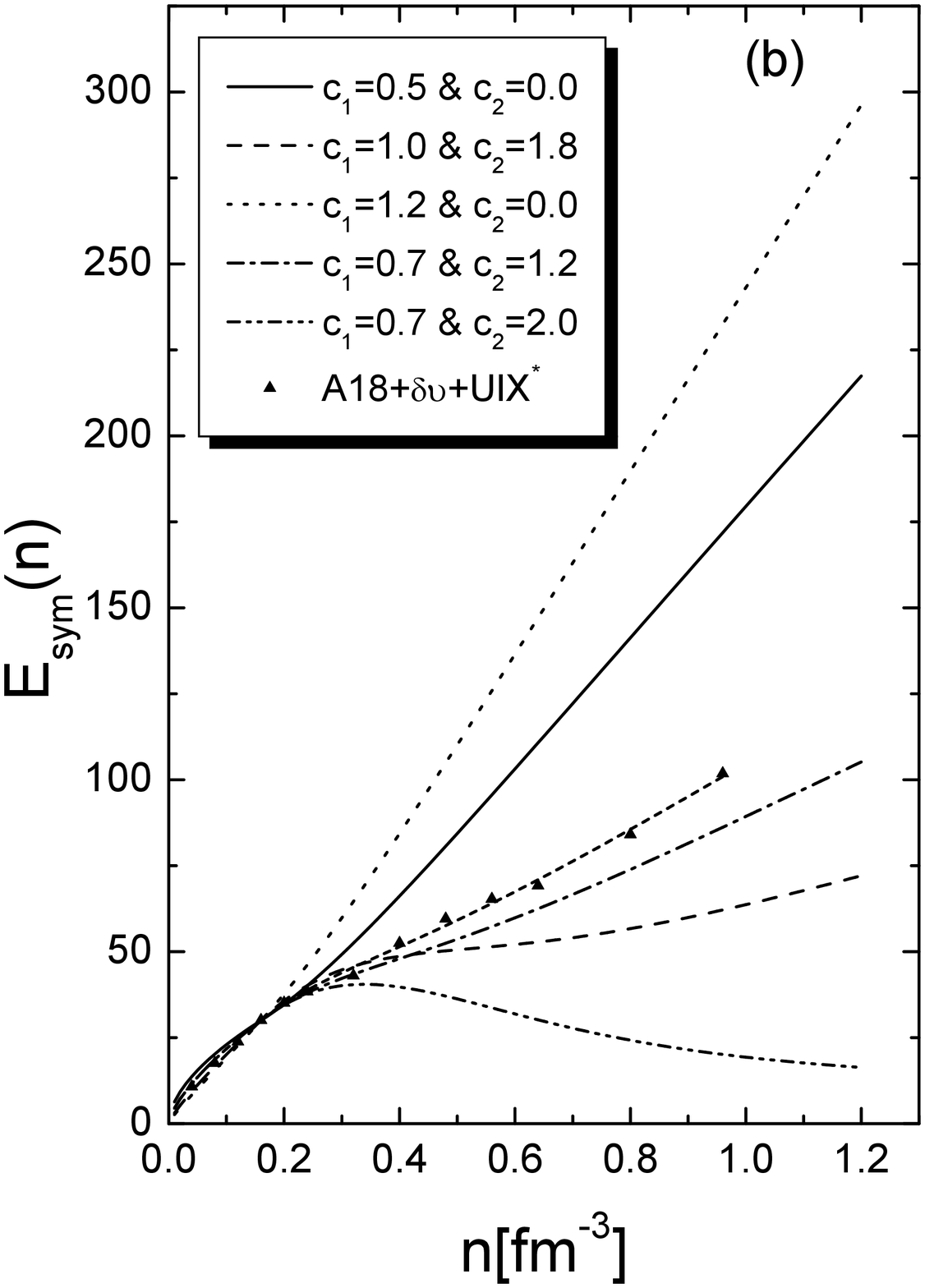}
 \caption{(a) $E_{sym}(n)$ for various values of the potential parameter
 $c$ of the function $F(u)$,
 given by the equation (\ref{fu-1}), versus the baryon density
 $n$. The case A18+$\delta$u+UIX$^{*}$ from Ref. \cite{Akmal98} is also included
 with the least-squares  fit curve which corresponds to the value $c=0.9$
 (b) $E_{sym}(n)$ for various values of  the potential parameters
 $c_1$ and $c_2$ of the function $F(u)$,
 given by the equation (\ref{FU-2}), versus the baryon density
 $n$. The case A18+$\delta$u+UIX$^{*}$ from Ref. \cite{Akmal98} is also included
 with the least-squares fit curve which corresponds to the values $c_1=0.77$ and $c_2=1.09$.}
 \label{totalrate0}
\end{figure}

\begin{figure}
\centering
\includegraphics[height=5.5cm,width=5.5cm]{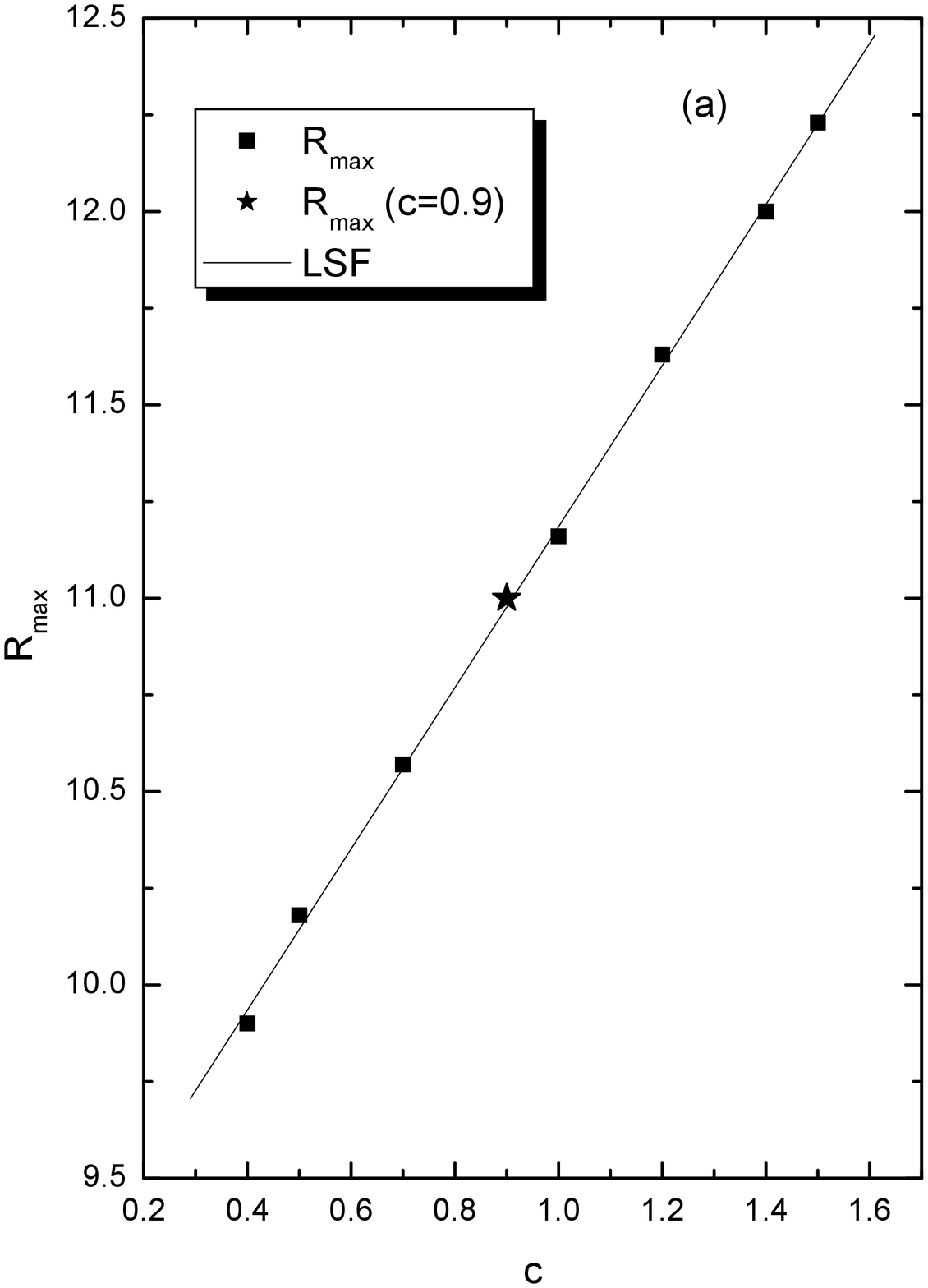}
\
 \includegraphics[height=5.5cm,width=5.5cm]{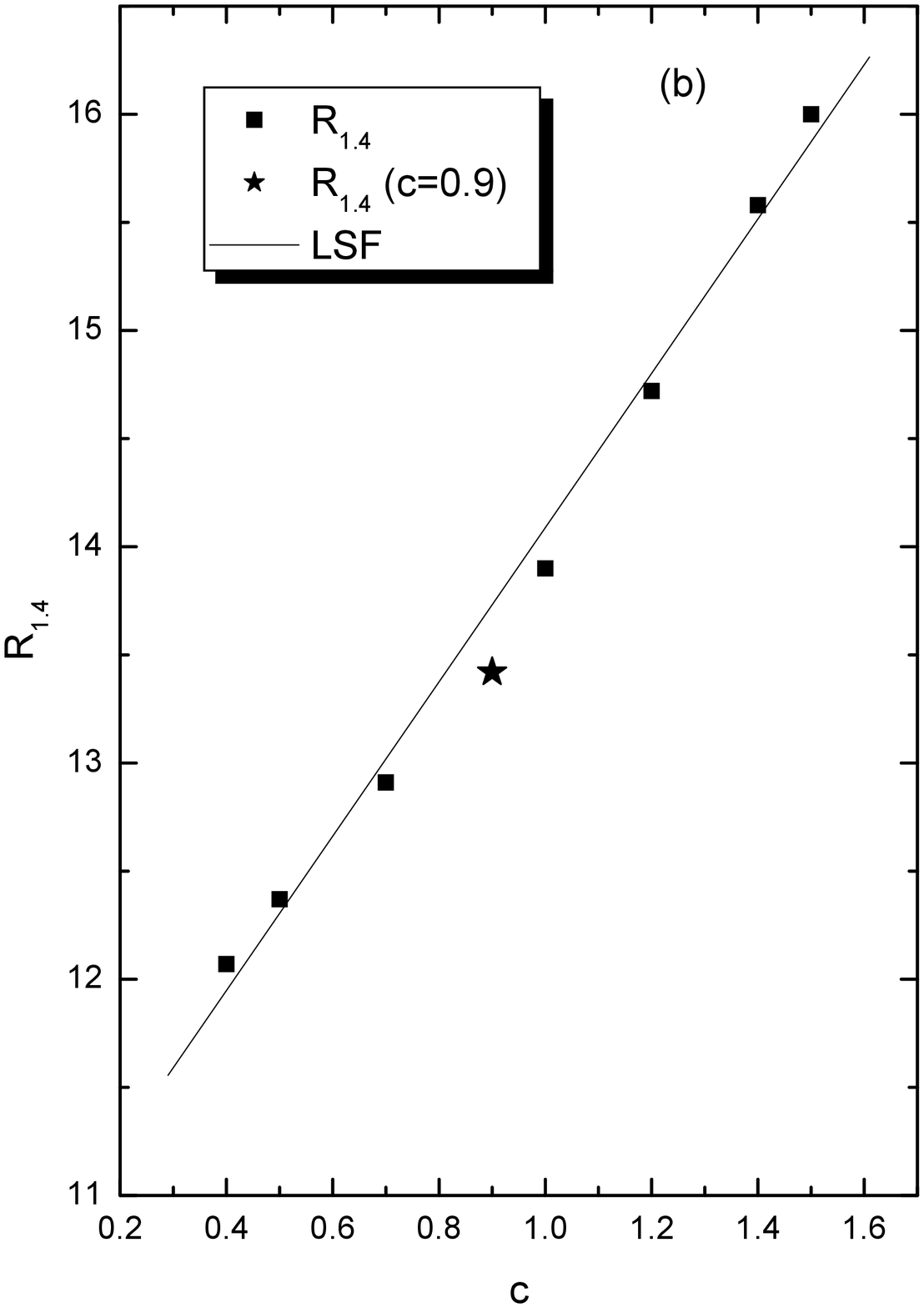}
 \
 \includegraphics[height=5.5cm,width=5.5cm]{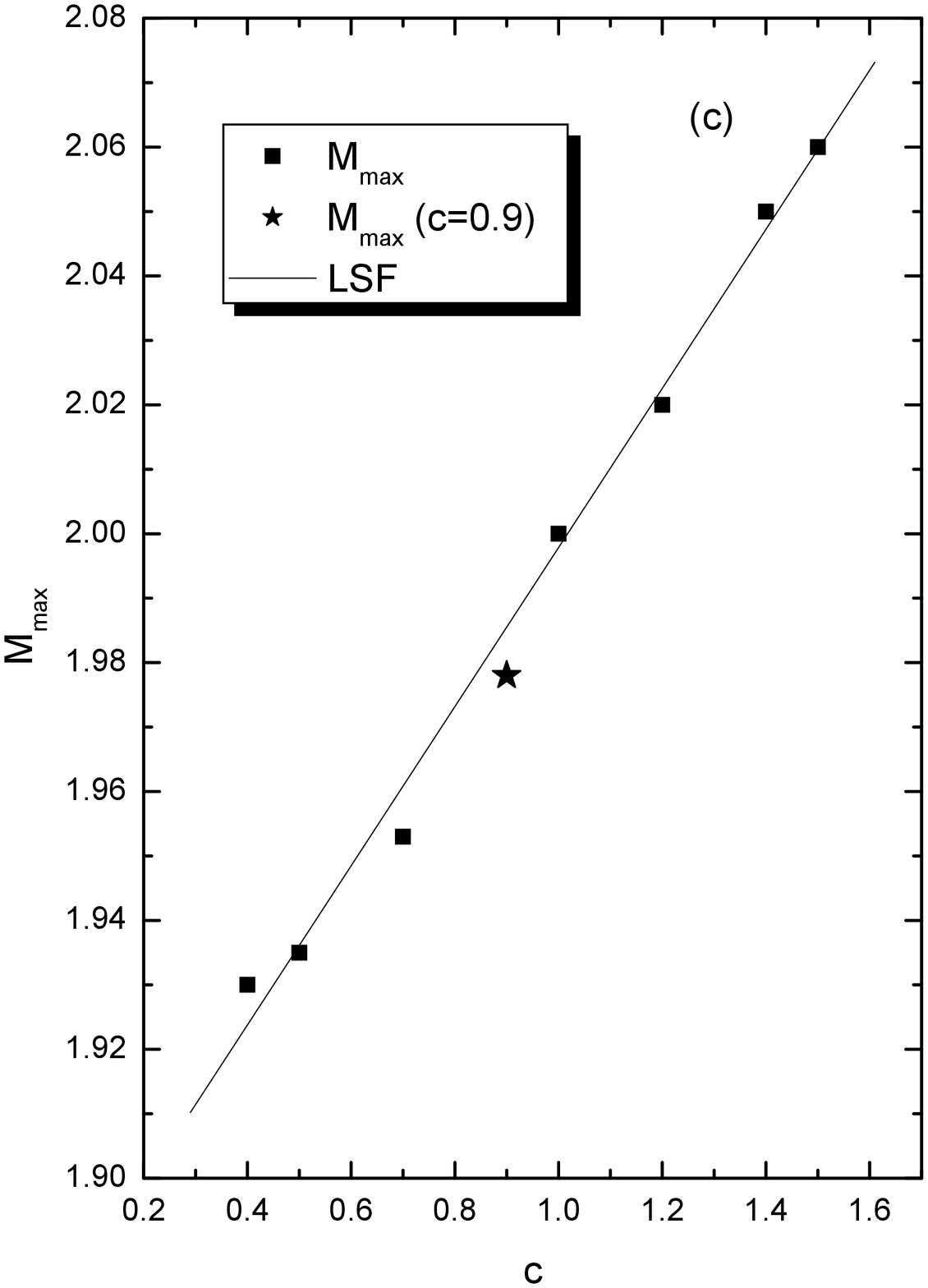}
\caption{(a) The radius $R_{max}$ as a function of the parameter
$c$. (b) The radius $R_{1.4}$ as a function of the parameter $c$.
(c) The maximum mass $M_{max}$ of the neutron star as a function
of the parameter c. The solid lines correspond to the least-squares
fit expressions (LSF) (\ref{Rmax-c-fit}), (\ref{R14-c-fit}) and
(\ref{Mmax-c-fit}) respectively. In all figures the star symbol corresponds to the case
 $A18+\delta u+UIX^{*}$ ( $c=0.9$).  } \label{}
\end{figure}
\begin{figure}
 \includegraphics[height=8.0cm,width=8.0cm]{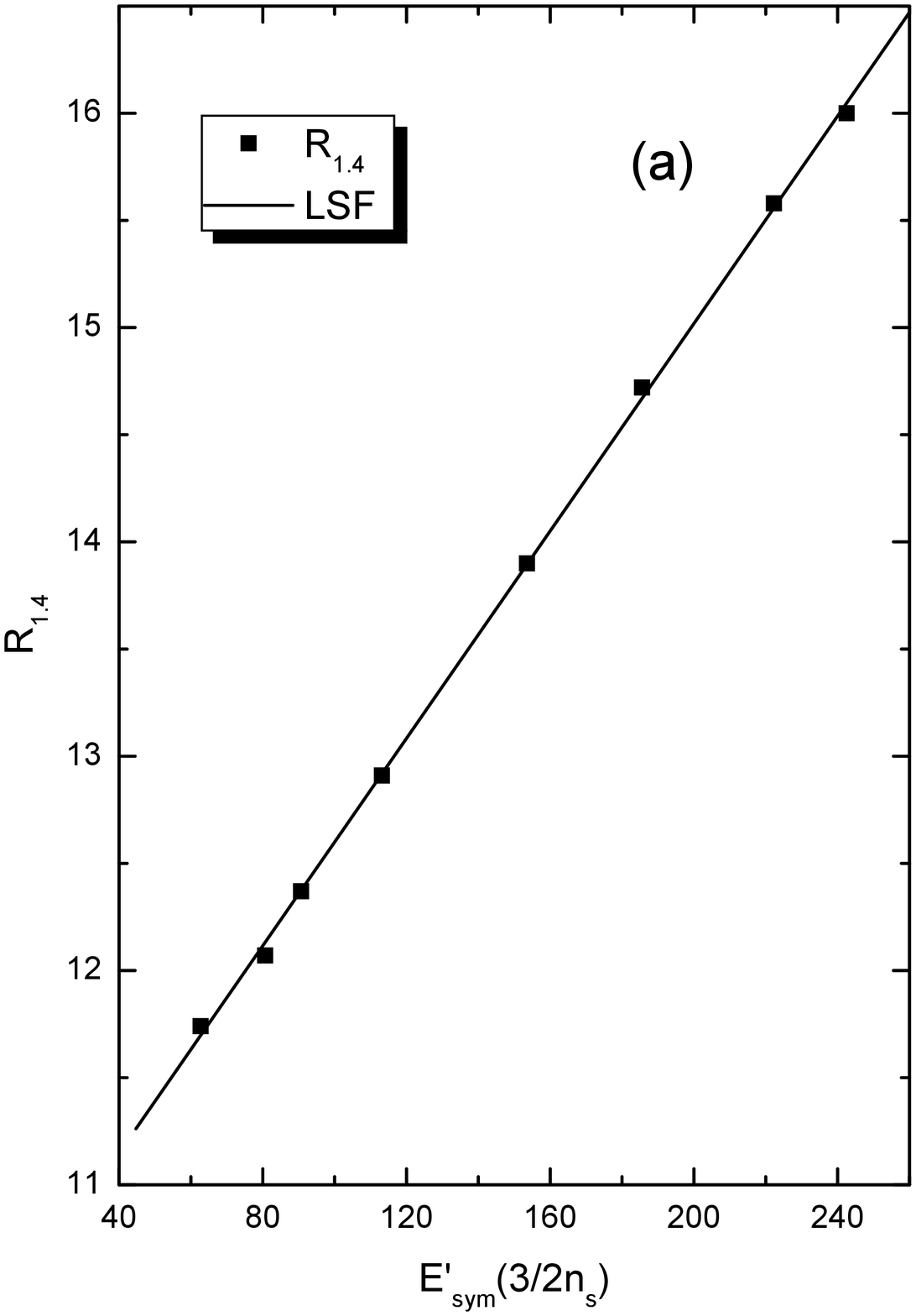}
\
 \includegraphics[height=8.0cm,width=8.0cm]{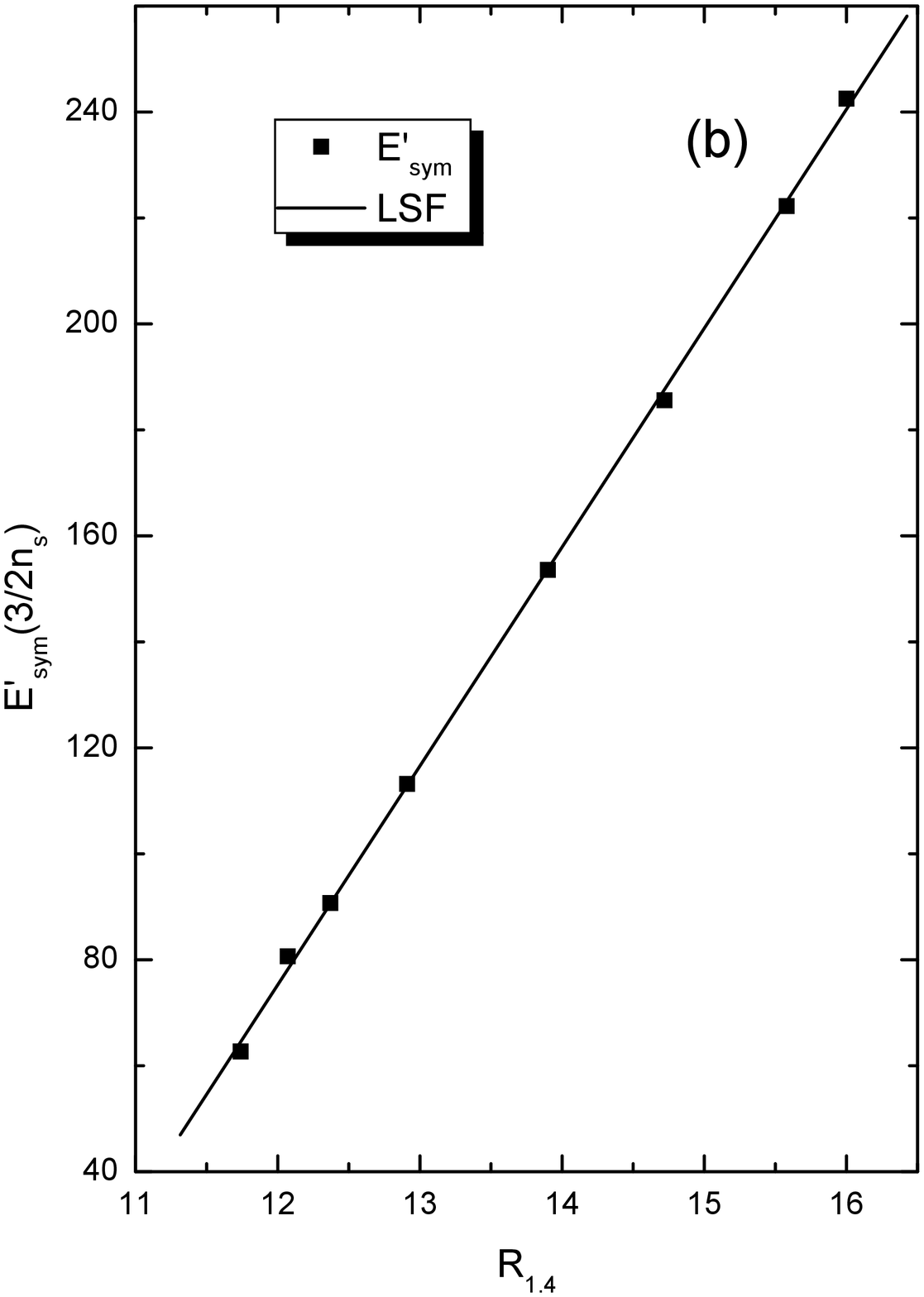}
 \caption{(a) The radius $R_{1.4}$ versus the derivative of the Symmetry
 Energy $E'_{sym}(3n_s/2)$. (b) $E'_{sym}(3n_s/2)$ versus  $R_{1.4}$.
  The solid lines correspond to the least-squares  fit (LSF) expressions
 (\ref{R14-Esym-fit}) and (\ref{Esym-R14-fit}) respectively. }
 \label{totalrate0}
\end{figure}
\begin{figure}
 \includegraphics[height=8.0cm,width=8.0cm]{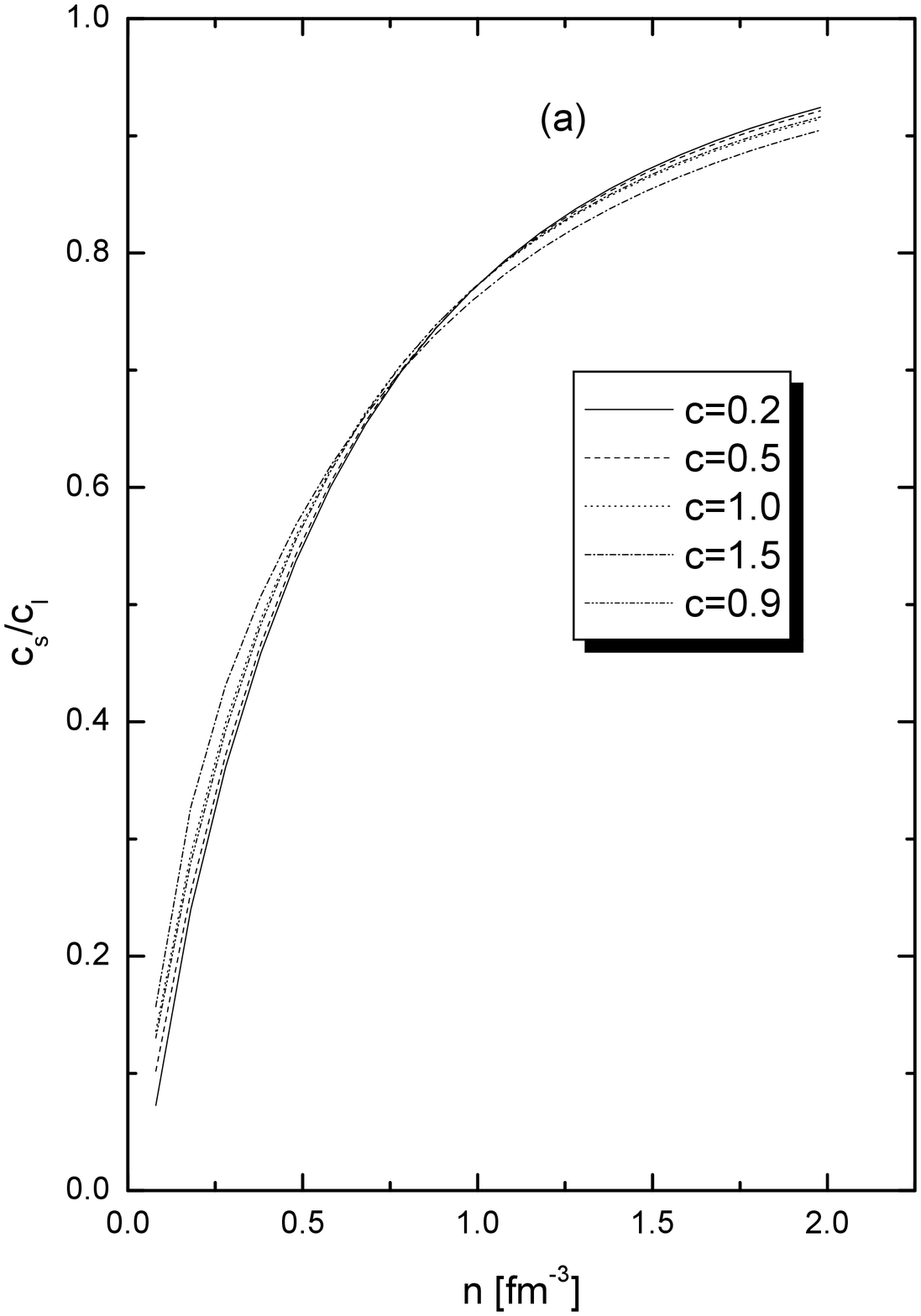}
\
 \includegraphics[height=8.0cm,width=8.0cm]{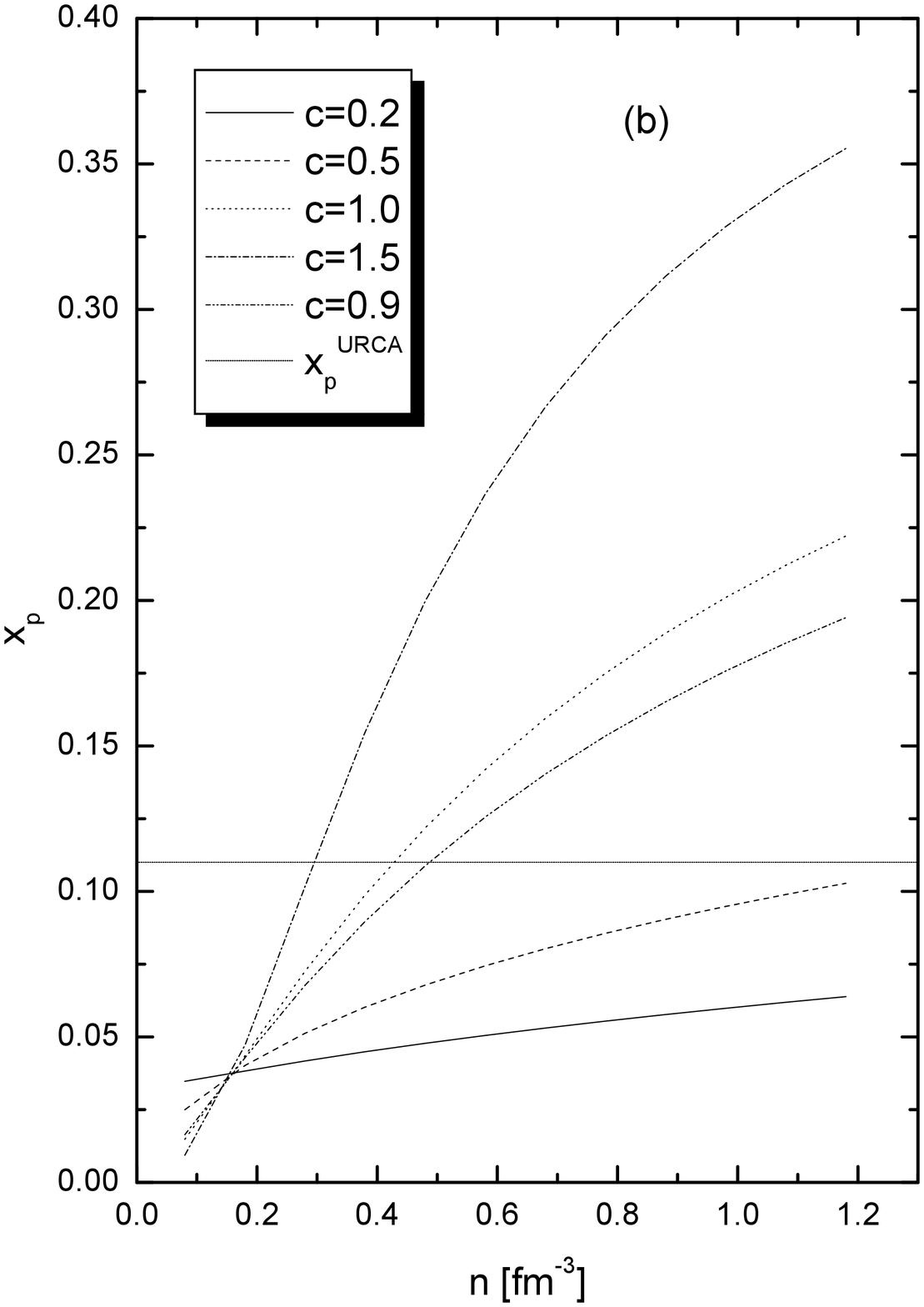}
 \caption{(a) The ratio $c_s/c_l$ versus the baryon density $n$ for
 various values of the potential parameter $c$. The line for $c=0.9$
 corresponds to the case $A18+\delta u+UIX^{*}$. (b) The proton fraction $x_p$
 versus the density $n$ for various values of the potential parameter $c$.
 The line for $c=0.9$
 corresponds to the case $A18+\delta u+UIX^{*}$.
 The short-dotted line
 shows the beginning the direct Urca process ($x_p=0.11$). }
 \label{totalrate0}
\end{figure}
\begin{figure}
 \includegraphics[height=8.0cm,width=8.0cm]{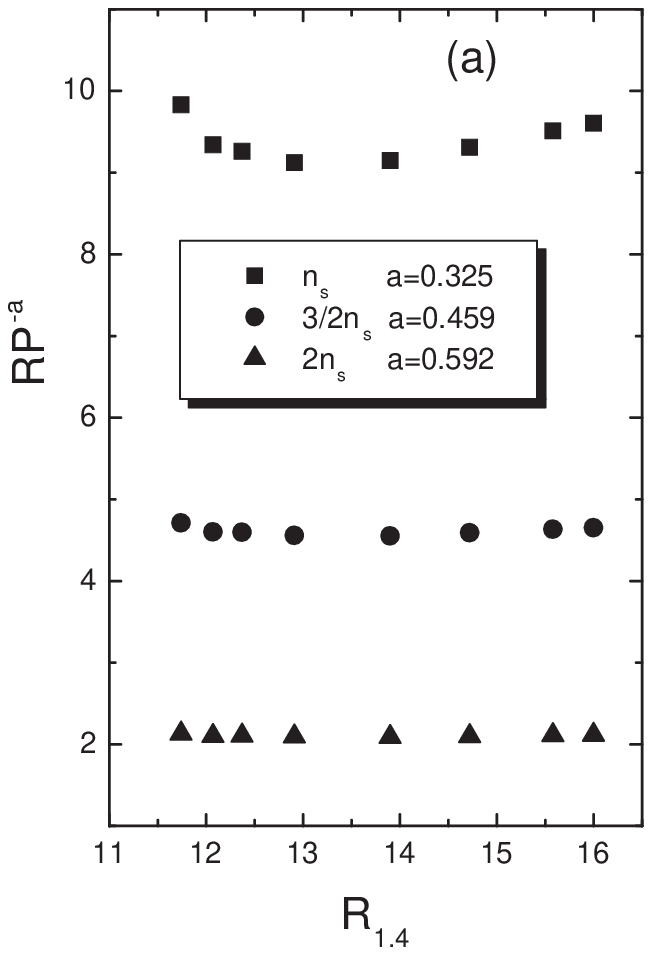}
\
 \includegraphics[height=8.0cm,width=8.0cm]{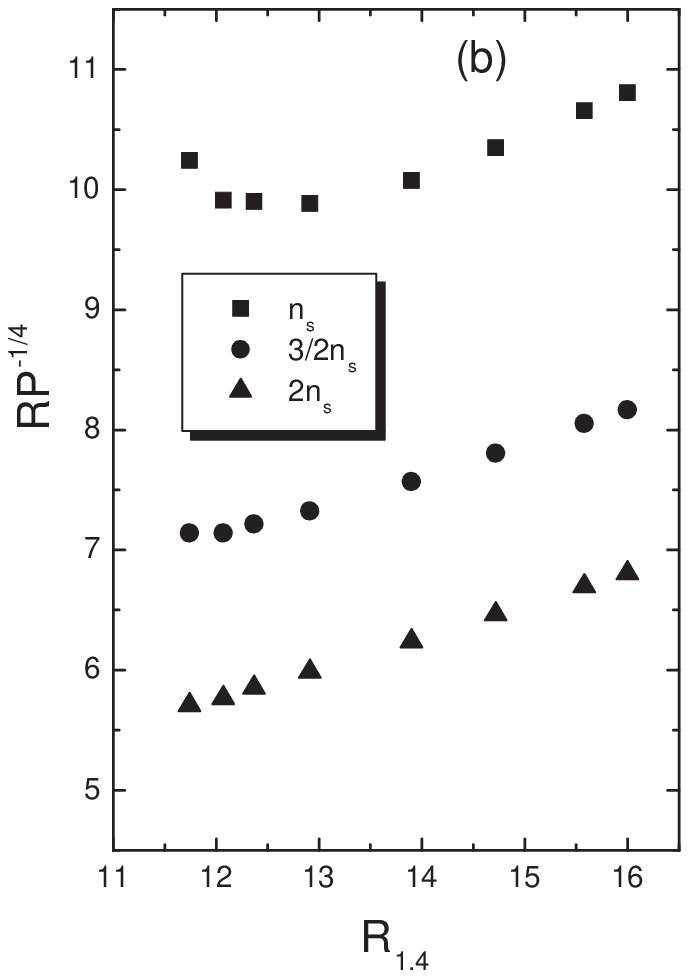}
 \caption{(a) The quantity $RP^{-a}$ as a function of the radius $R_{1.4}$
 for pressure determined at $n=n_s$, $n=3n_s/2$ and $n=2n_s$. For each density,
 the least-squares fit value for the exponent
 $a$ is indicated. (b) The quantity $RP^{-1/4}$ as
 a function of the radius $R_{1.4}$
 for pressure determined at $n=n_s$, $n=3n_s/2$ and $n=2n_s$.}
 \label{totalrate0}
\end{figure}
\begin{figure}
 \includegraphics[height=8.0cm,width=8.0cm]{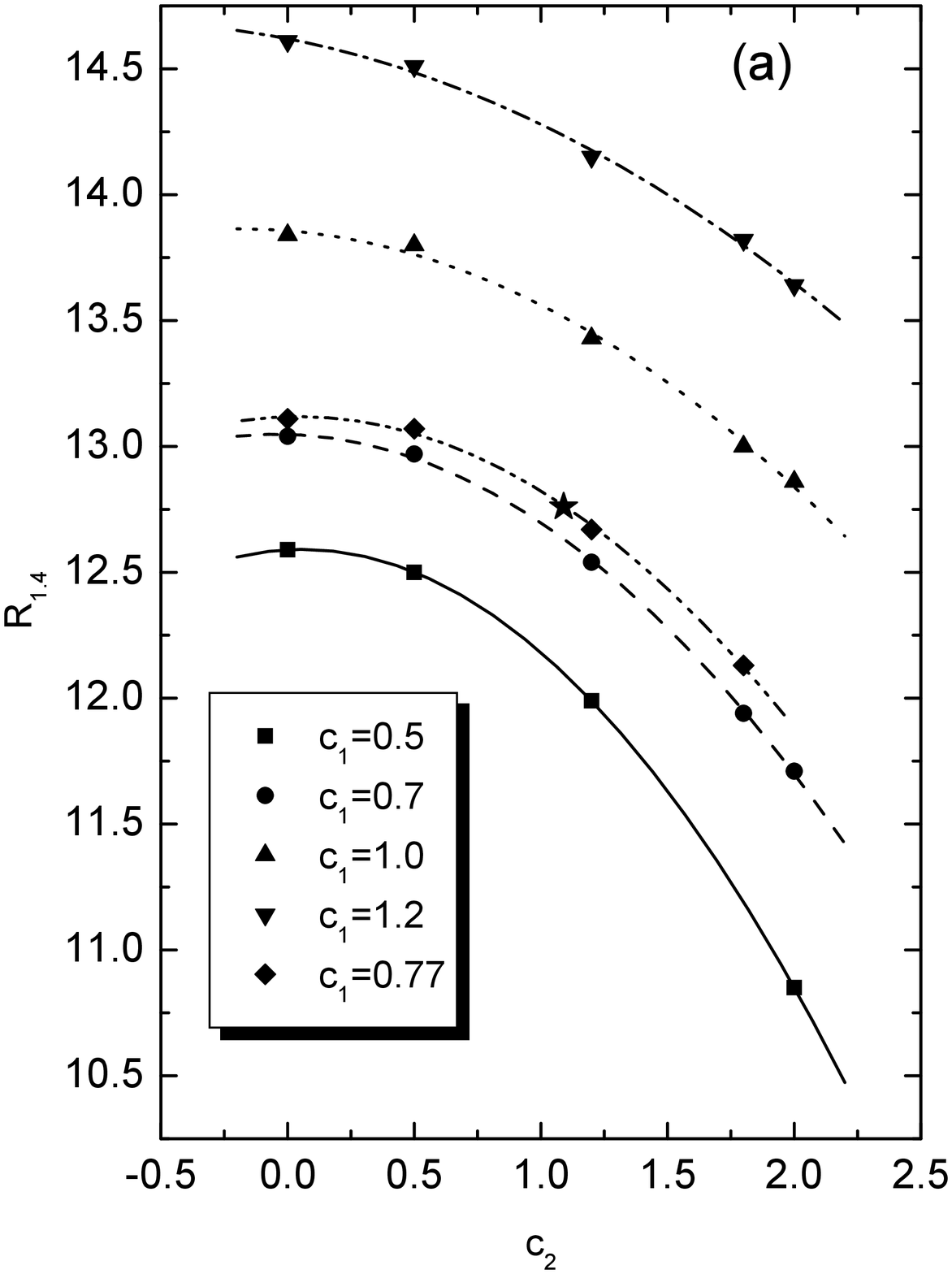}
\
 \includegraphics[height=8.0cm,width=8.0cm]{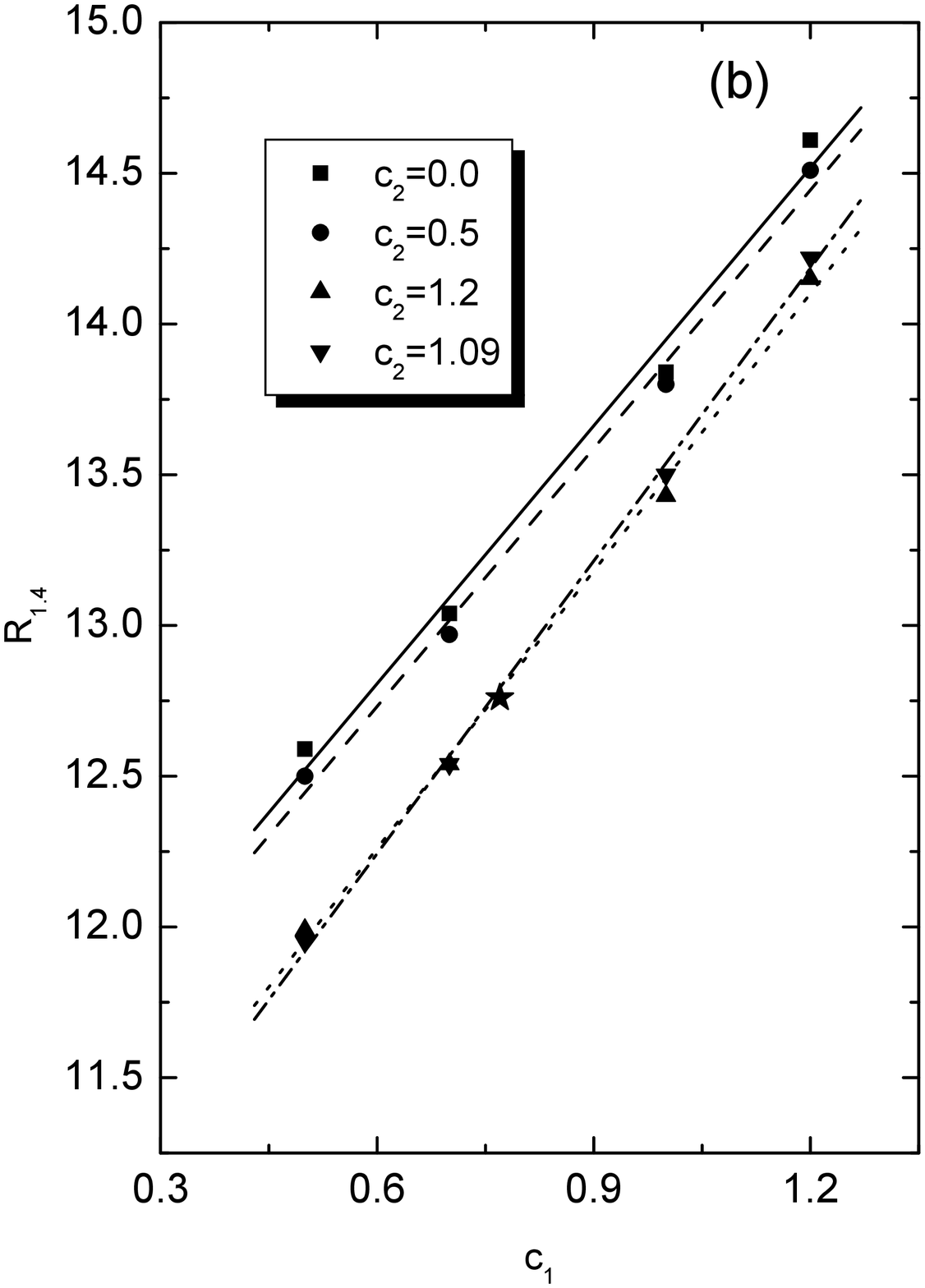}
 \\
 \includegraphics[height=8.0cm,width=8.0cm]{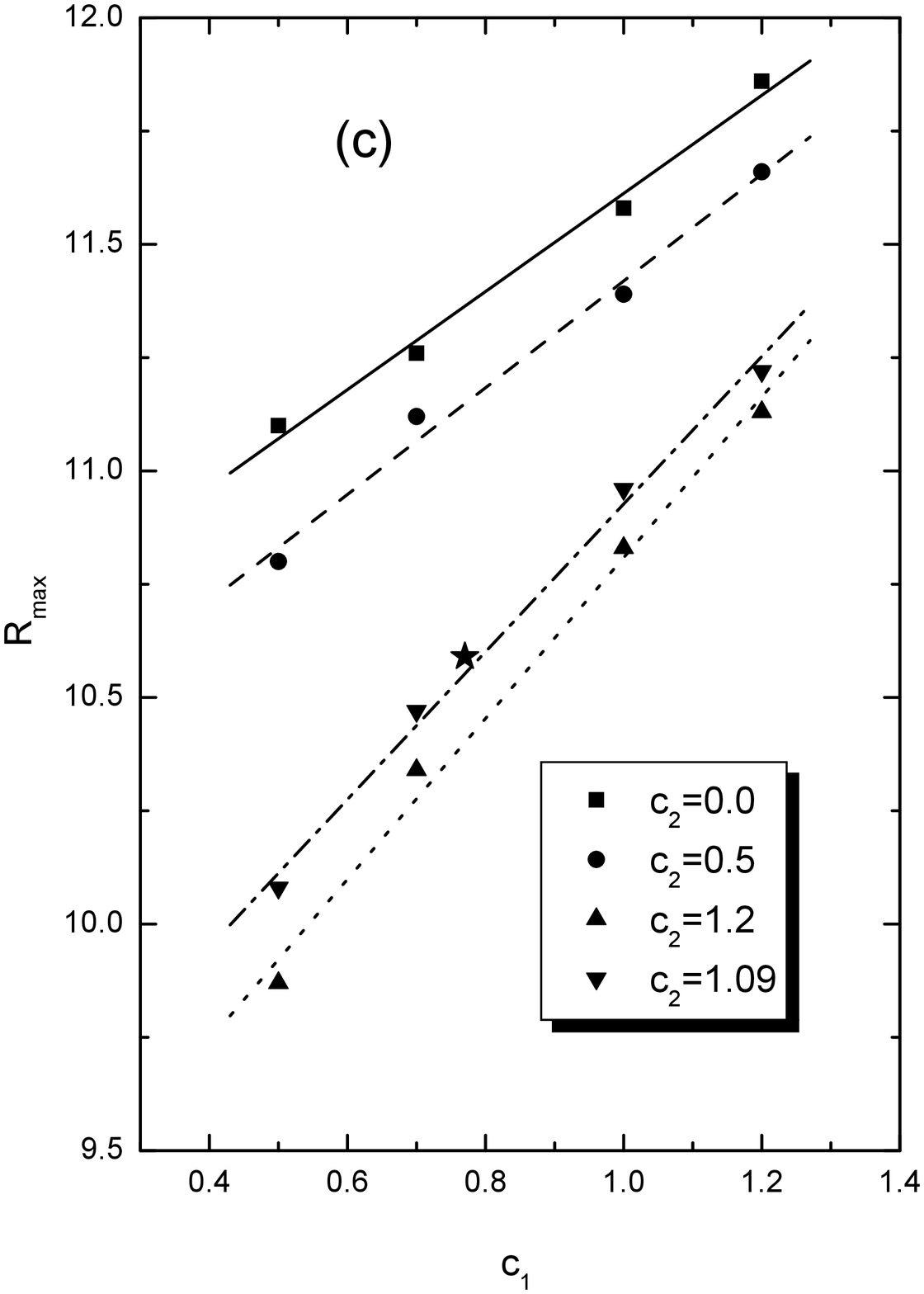}
\
 \includegraphics[height=8.0cm,width=8.0cm]{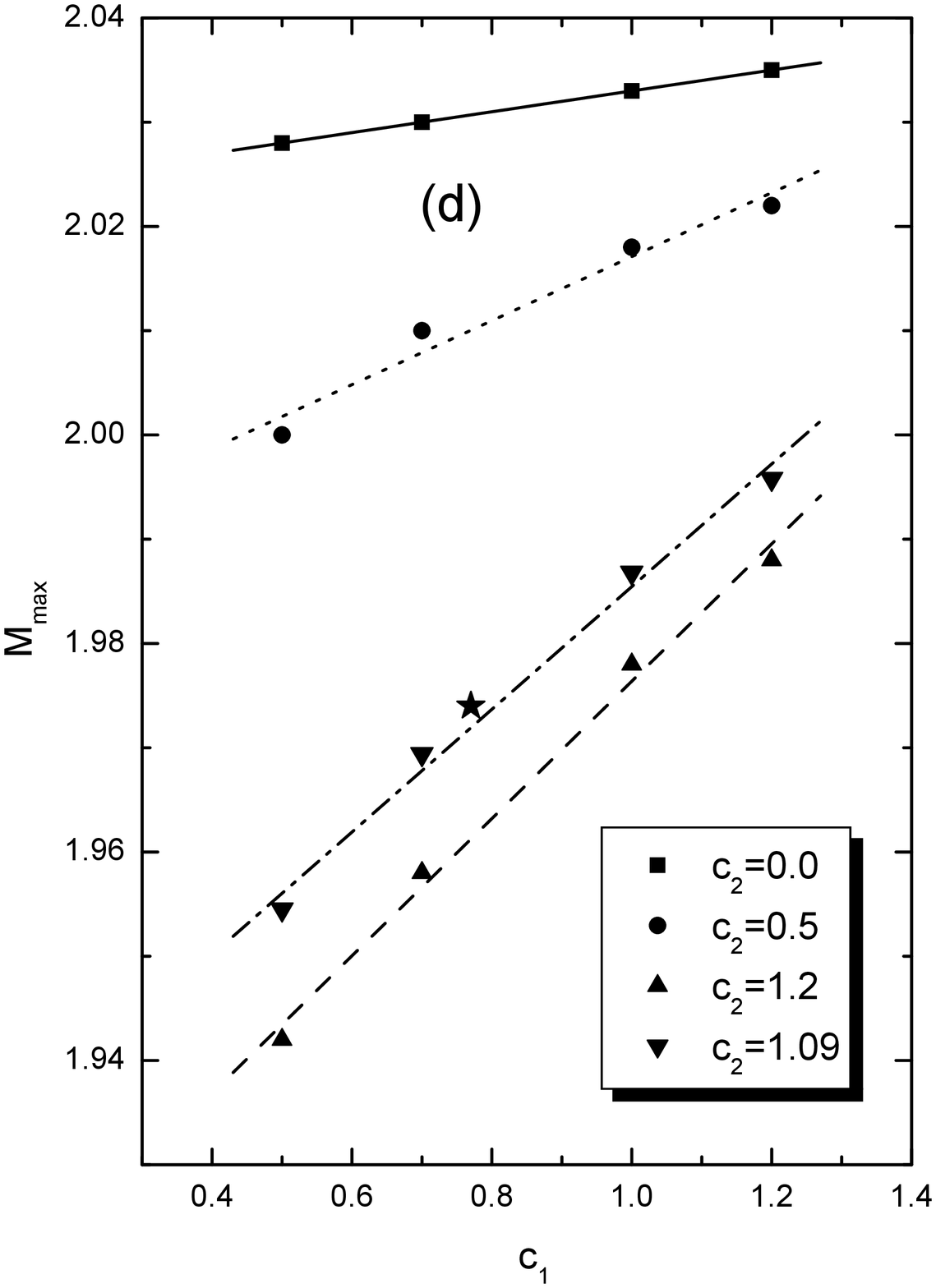}
 \caption{(a) The radius $R_{1.4}$ as a function of the
 second potential parameter $c_2$ for various values of the first
 potential parameter $c_1$. The lines correspond to the least-squares fit
 expressions (\ref{R-c1-c2-a}) .
 (b) The radius $R_{1.4}$ as a function of the
 first potential parameter $c_1$ for various values of the second
 potential parameter $c_2$. The lines correspond to the least-squares fit
 expressions (\ref{R-c1-c2}).
 (c) The radius $R_{max}$  as a function of the
 first potential parameter $c_1$ for various values of the second
 potential parameter $c_1$. The lines correspond to the least-squares fit
 expressions.
 (d) The maximum mass $M_{max}$ for various values of the second
 potential parameter $c_1$. The lines correspond to the least-squares fit
 expressions. In all figures the star symbol corresponds to the case
 $A18+\delta u+UIX^{*}$ ( $c_1=0.77$ and $c_2=1.09$)   }
 \label{totalrate0}
\end{figure}
\begin{figure}
 \includegraphics[height=8.0cm,width=8.0cm]{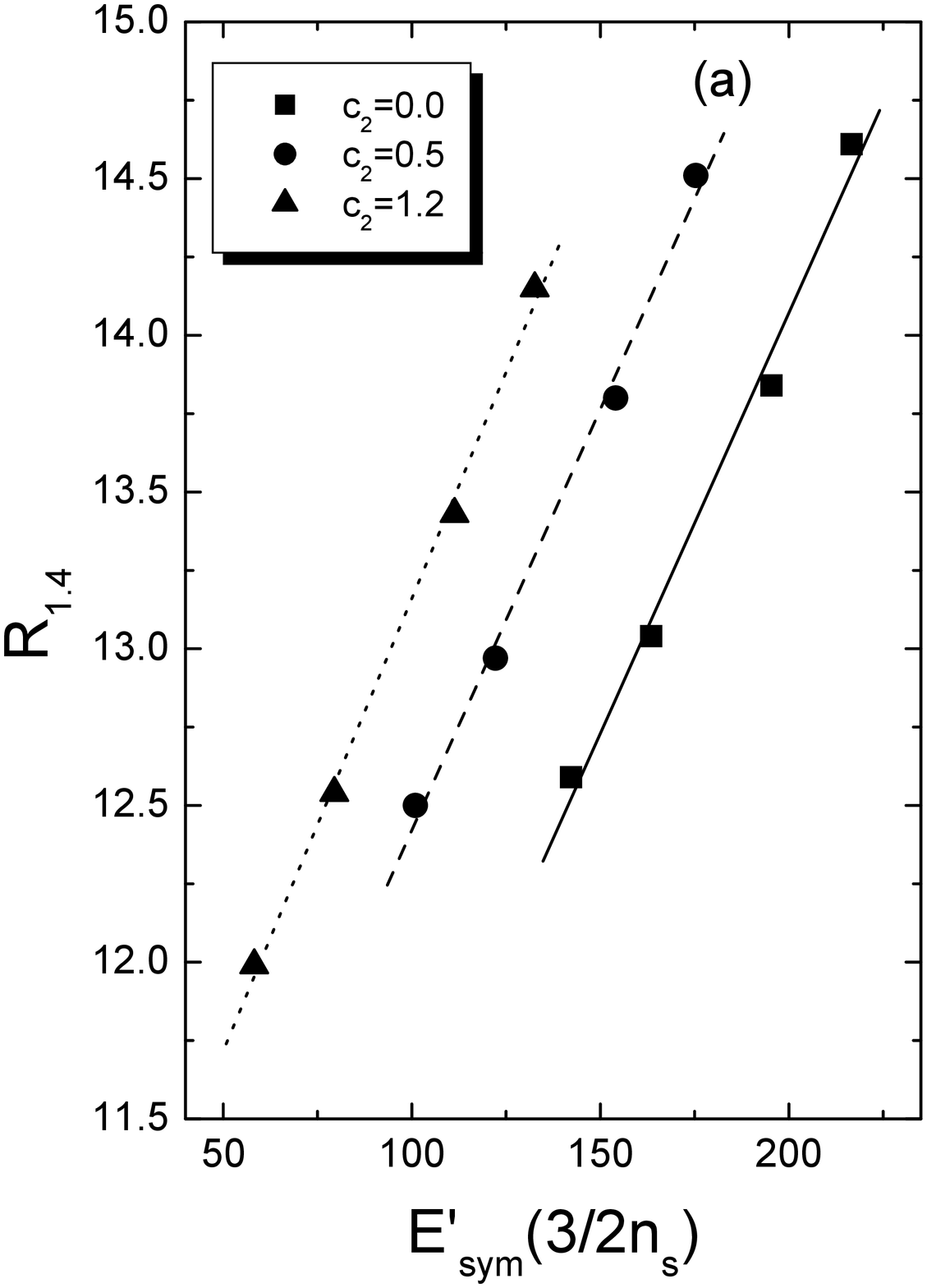}
\
 \includegraphics[height=8.0cm,width=8.0cm]{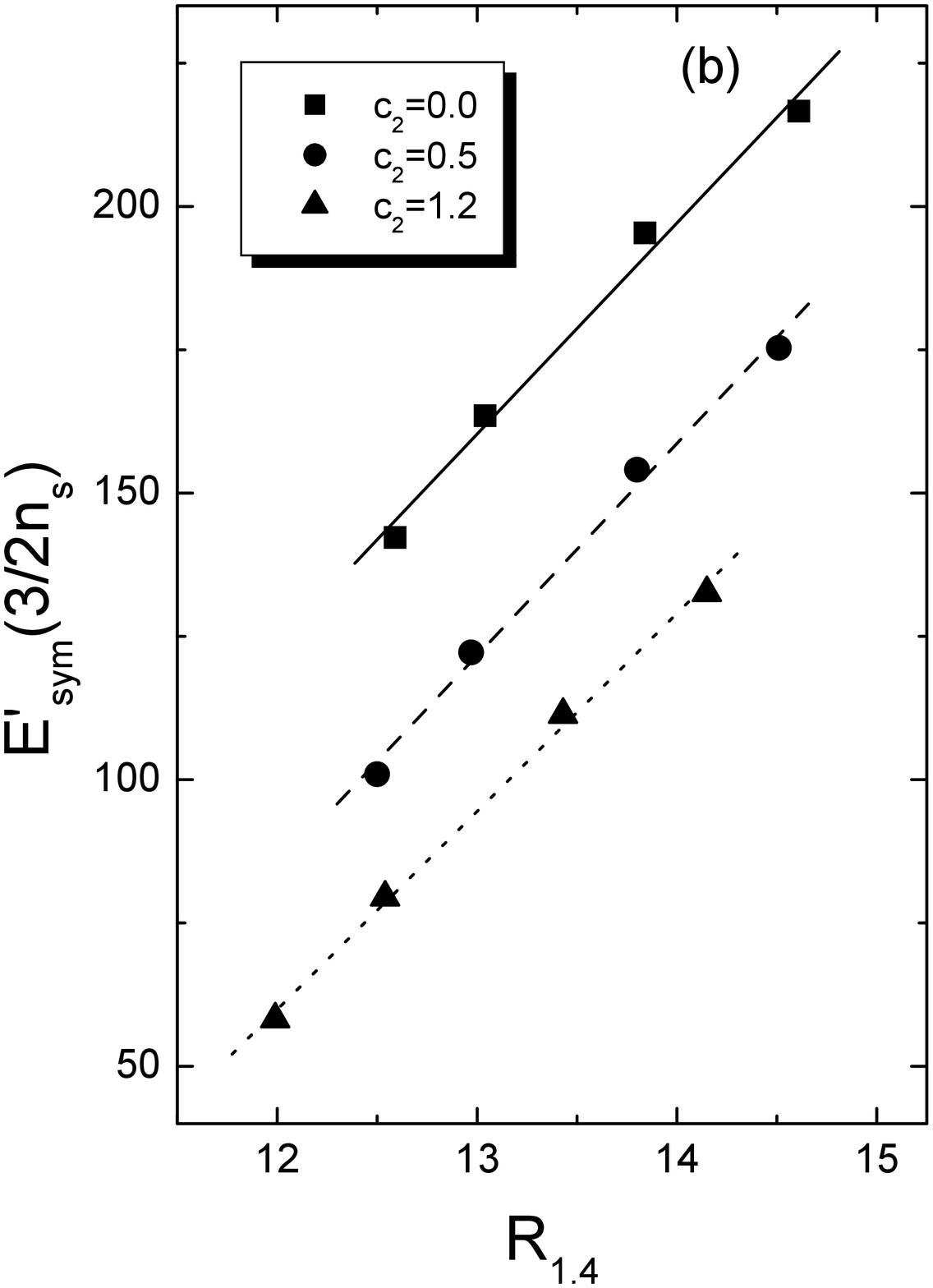}
 \caption{(a) The radius $R_{1.4}$ versus the derivative of the symmetry
 energy $E'_{sym}(3n_s/2)$ for various values of the potential parameter $c_2$.
 (b) $E'_{sym}(3n_s/2)$ versus  $R_{1.4}$. The lines correspond to the least-squares fit expressions
 (\ref{R-Esym-0}), (\ref{R-Esym-05}) and (\ref{R-Esym-12}) respectively. }
 \label{totalrate0}
\end{figure}
\begin{figure}
 \includegraphics[height=8.0cm,width=8.0cm]{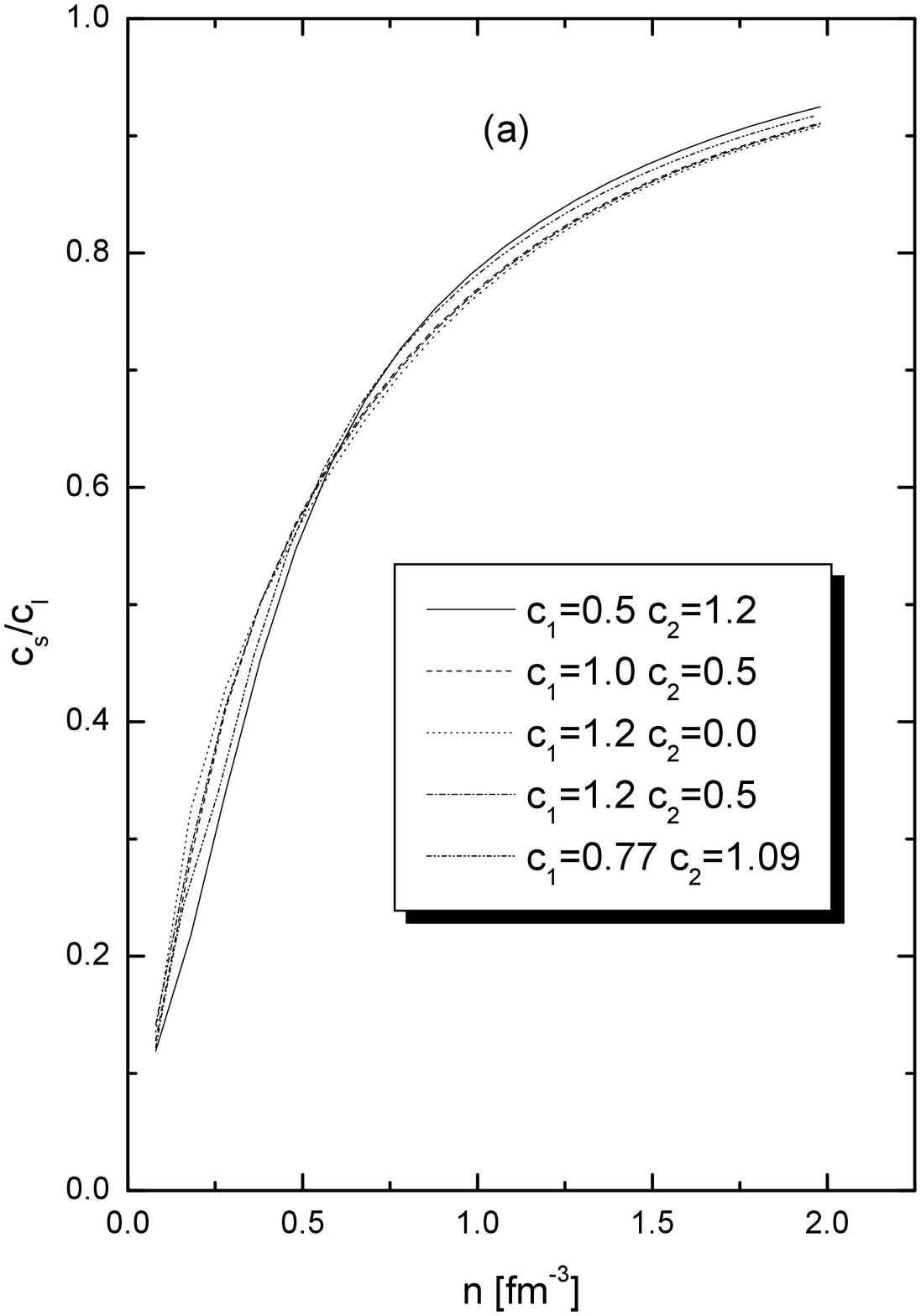}
\
 \includegraphics[height=8.0cm,width=8.0cm]{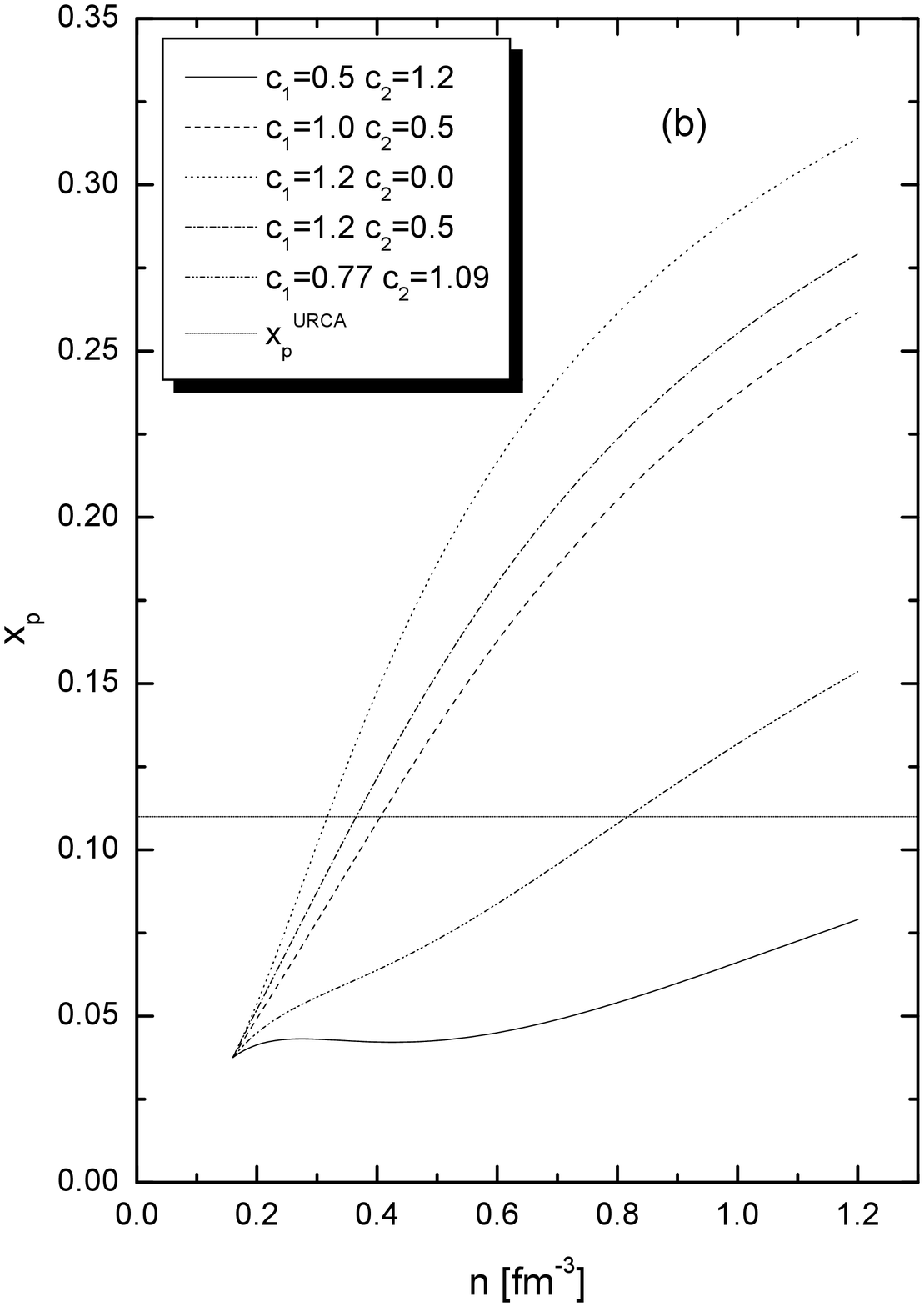}
 \caption{(a) The ratio $c_s/c_l$ versus the baryon density $n$ for
 various values of the potential parameters $c_1$ and $c_2$.
 The line for $c_1=0.77$ and $c_2=1.09$
 corresponds to the case $A18+\delta u+UIX^{*}$.
 (b) The proton fraction $x_p$
 versus the density $n$ for various values of the potential parameters $c_1$ and $c_2$.
 The line for $c_1=0.77$ and $c_2=1.09$
 corresponds to the case $A18+\delta u+UIX^{*}$.
 The short-dotted line shows the beginning of the direct
 Urca process ($x_p=0.11$). }
 \label{totalrate0}
\end{figure}
\begin{figure}
 \includegraphics[height=8.5cm,width=5.8cm]{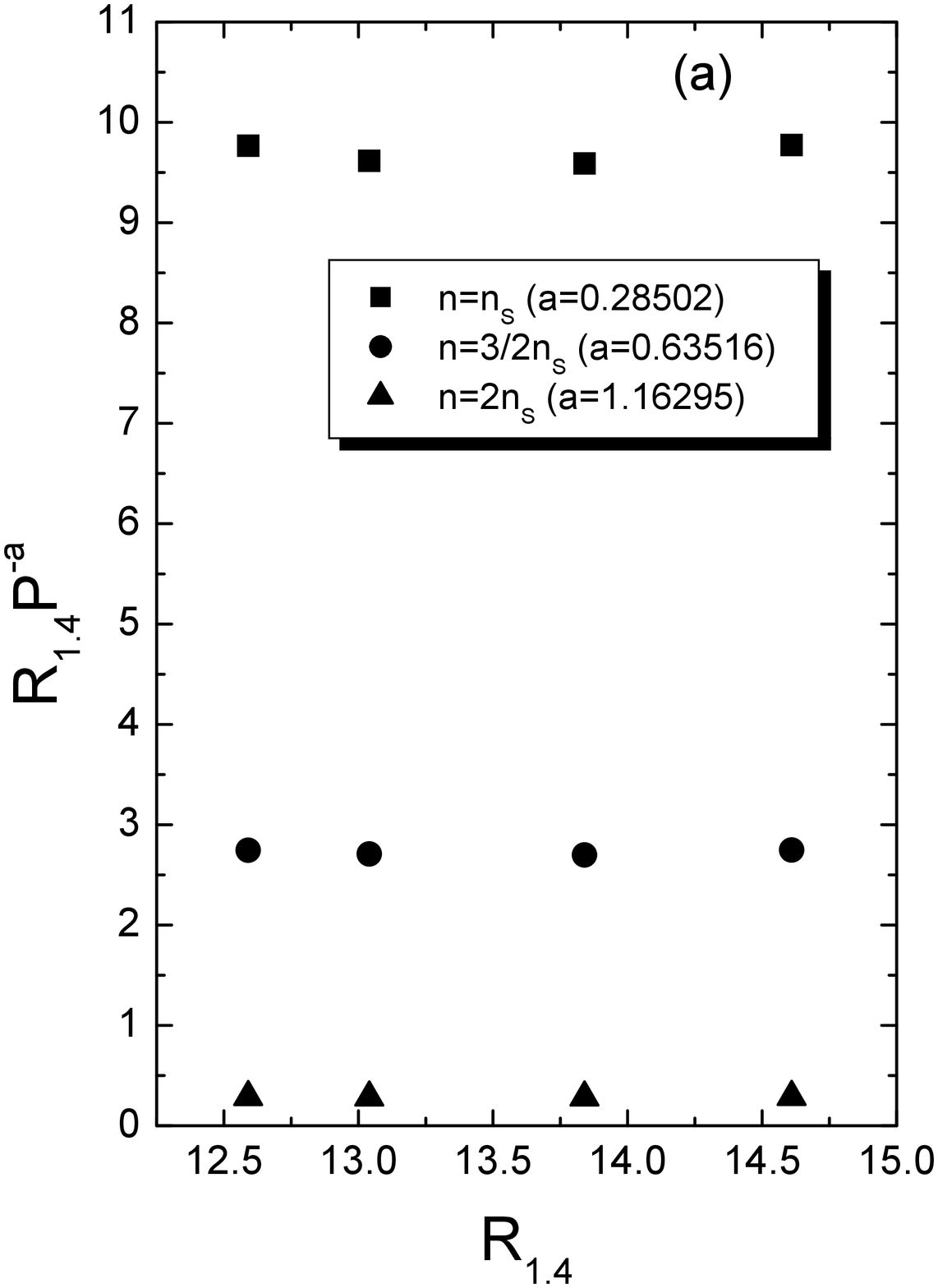}
\
 \includegraphics[height=8.5cm,width=5.8cm]{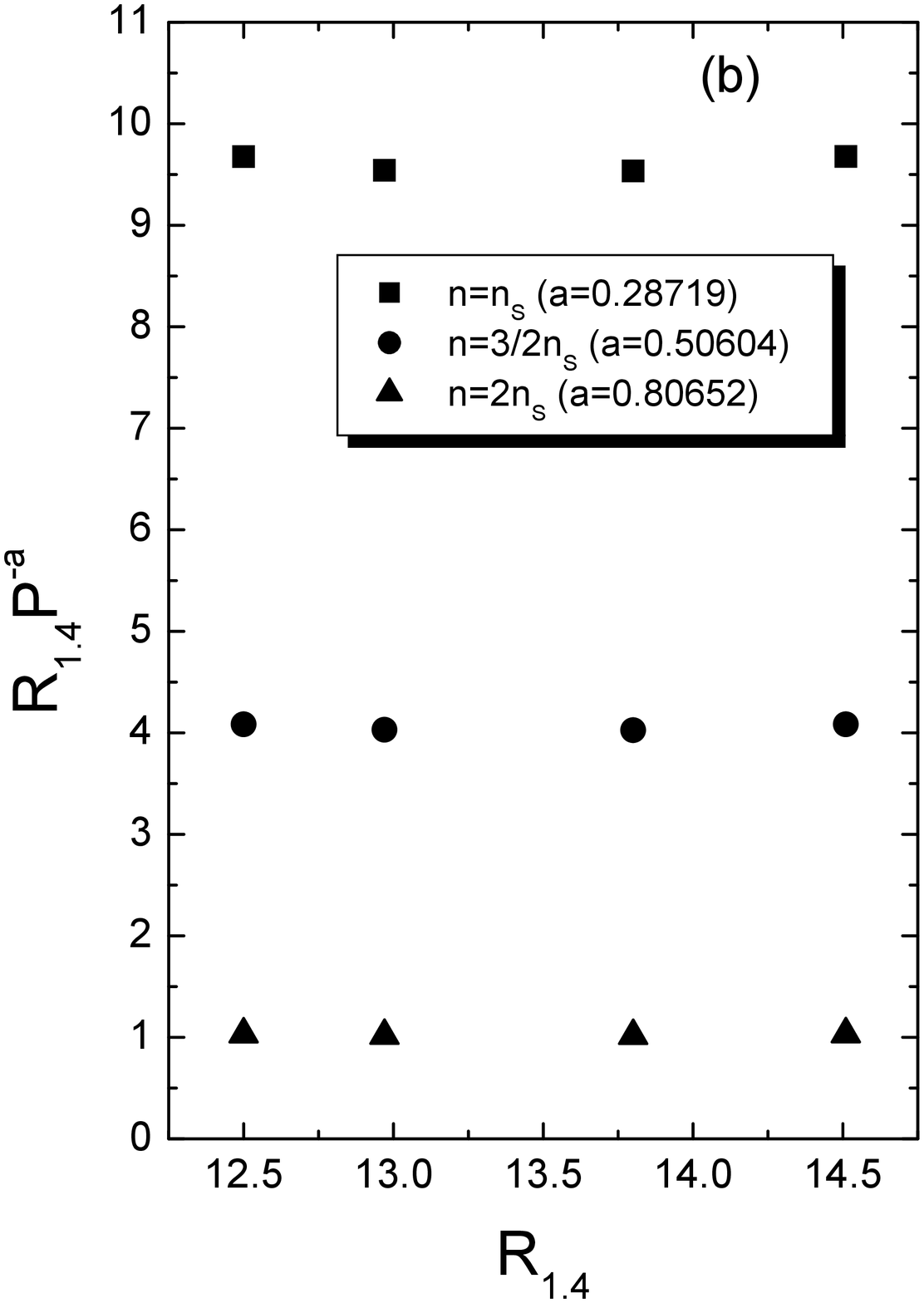}
 \
 \includegraphics[height=8.5cm,width=5.8cm]{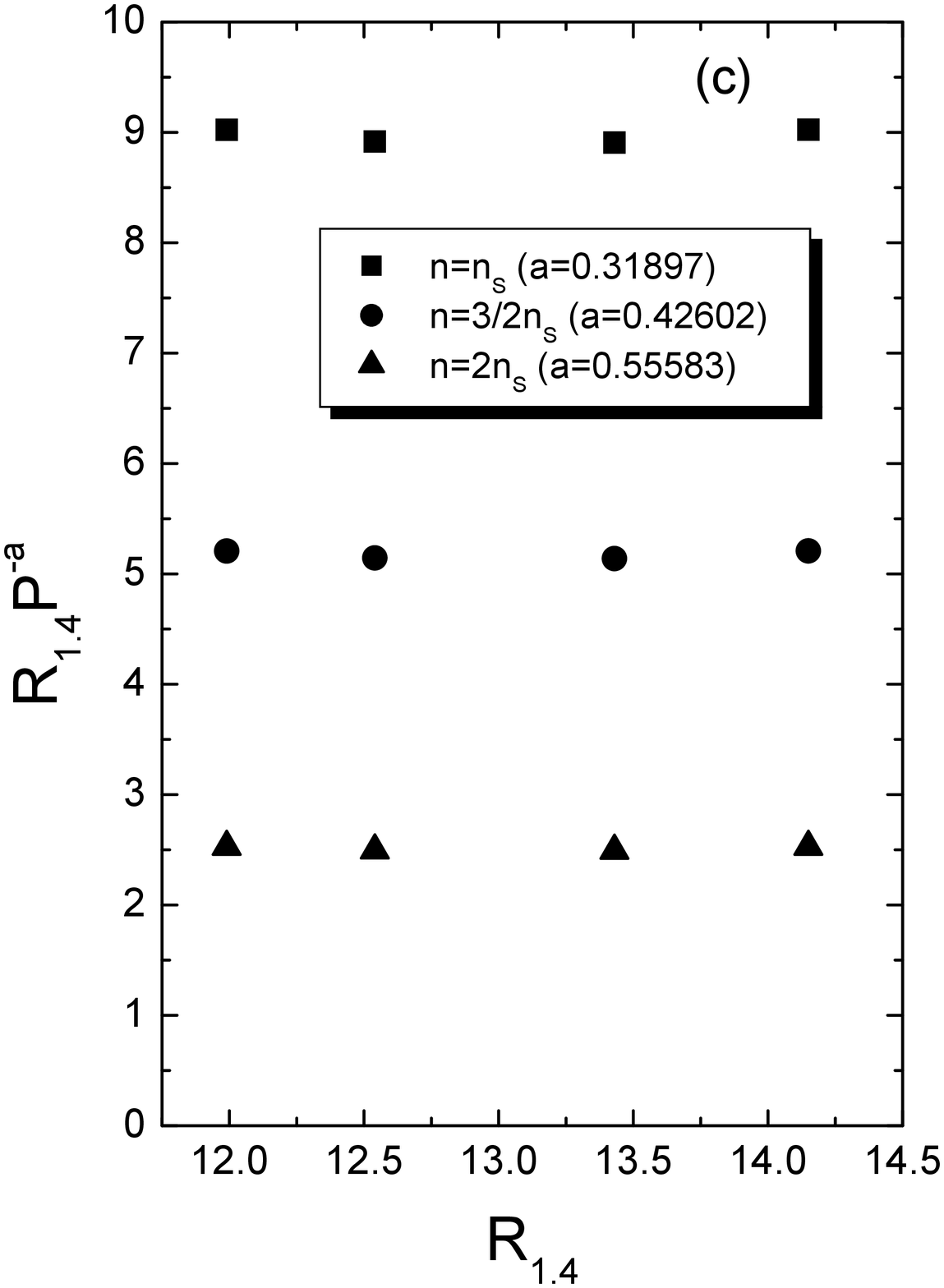}
 \caption{(a) The quantity $RP^{-a}$ as a function of the radius $R_{1.4}$
for pressure determined at $n=n_s$, $n=3n_s/2$ and $n=2n_s$ for
$c_2=0$. (b) The same as before for $c_2=0.5$. (c) The same as before for $c_2=1.2$. For each density,
 the least-squares fit value for the exponent a is indicated.  }
 \label{totalrate0}
\end{figure}
\begin{figure}
 \includegraphics[height=8.5cm,width=5.8cm]{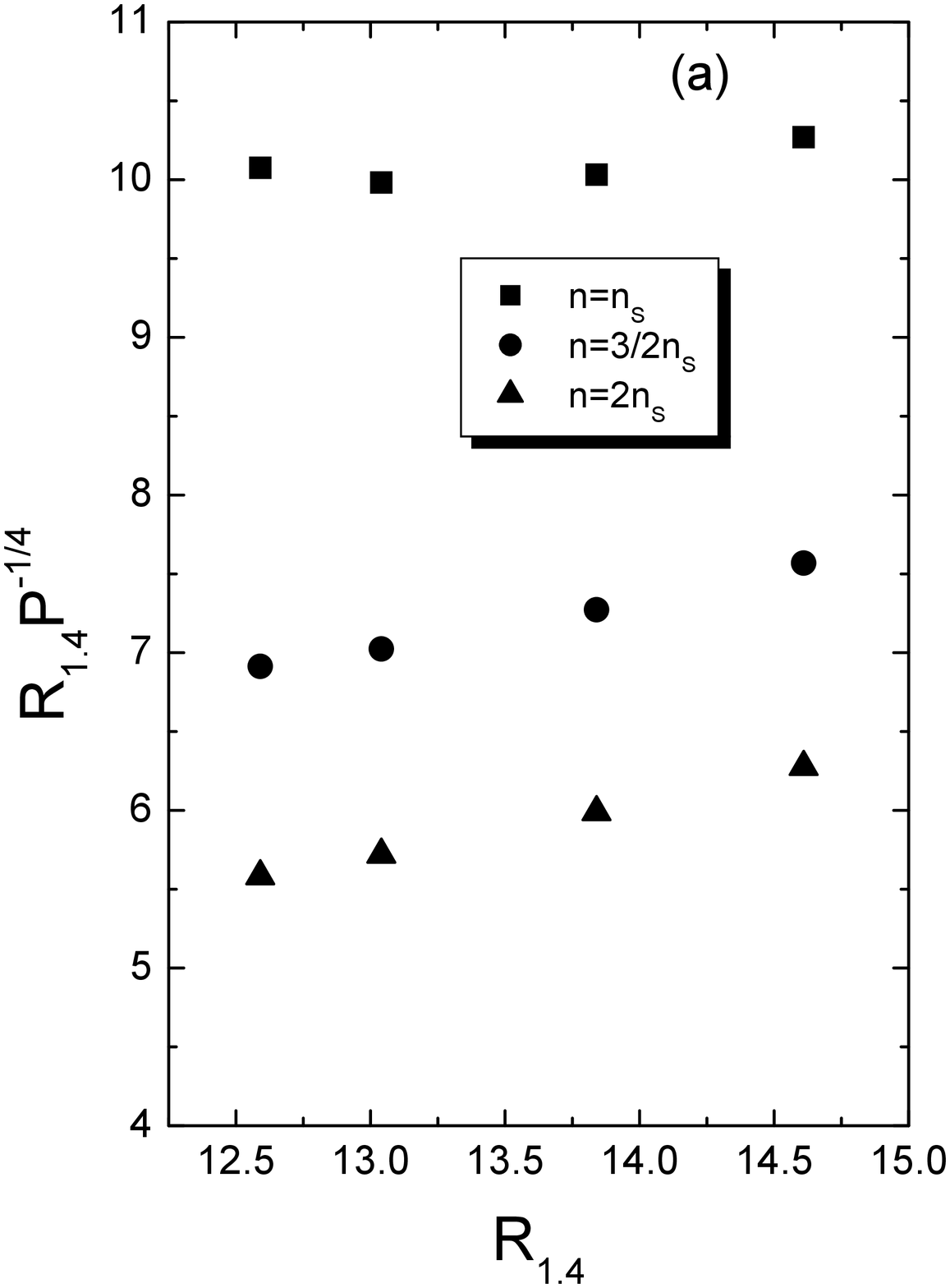}
\
 \includegraphics[height=8.5cm,width=5.8cm]{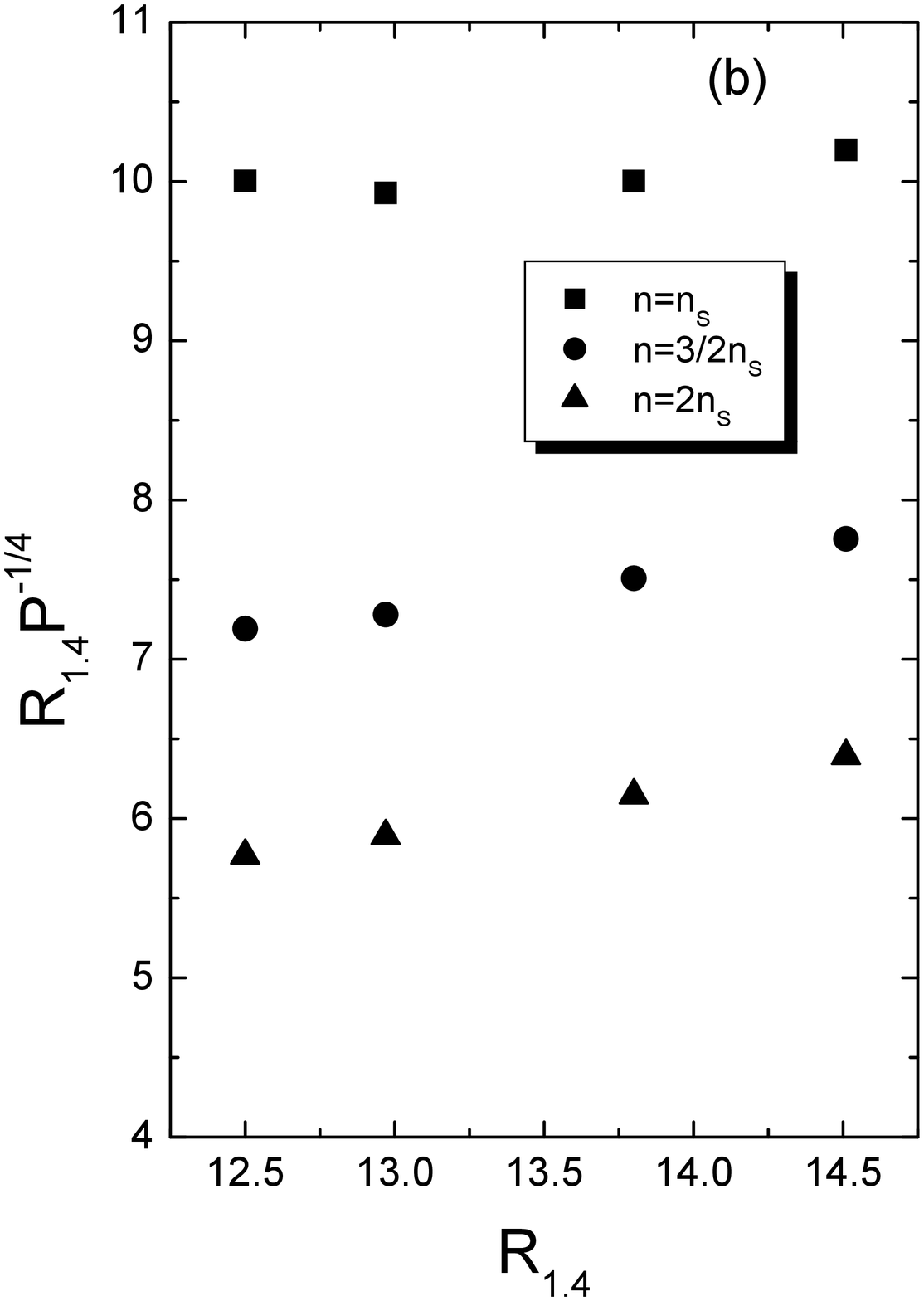}
 \
 \includegraphics[height=8.5cm,width=5.8cm]{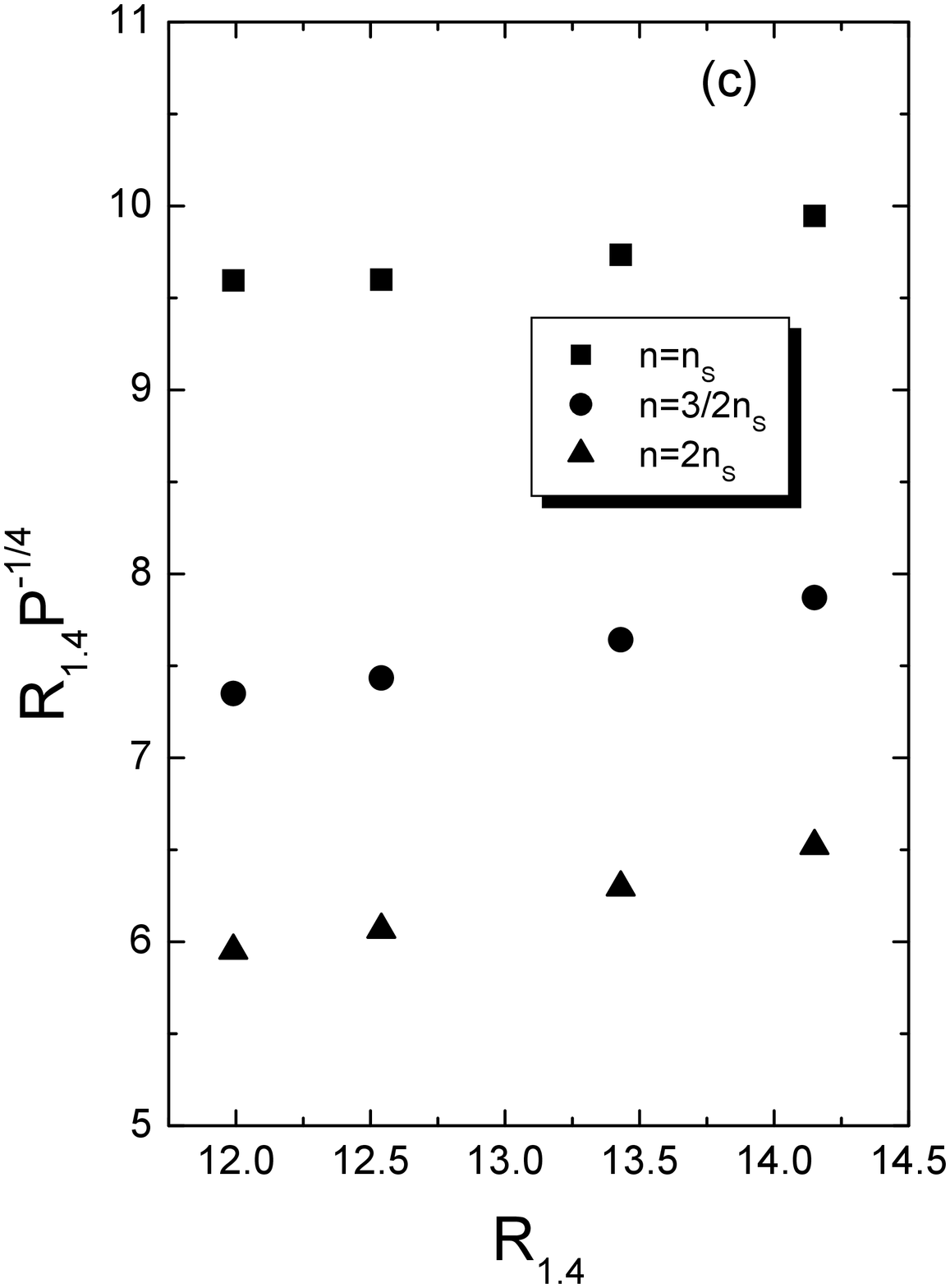}
 \caption{ (a) The quantity $RP^{-1/4}$ as a function of the radius $R_{1.4}$
for pressure determined at $n=n_s$, $n=3n_s/2$ and $n=2n_s$ for
$c_2=0$. (b) The same as before for $c_2=0.5$. (c) The same as before for $c_2=1.2$.}
 \label{totalrate0}
\end{figure}

\end{document}